\documentclass[12pt]{iopart}
\usepackage{iopams}
\usepackage{upgreek}
\usepackage{graphicx}
\usepackage[usenames,dvipsnames]{xcolor}
\usepackage{hyperref}
\usepackage{siunitx}
\usepackage[T1]{fontenc}
\usepackage[compress,numbers,square,comma]{natbib}
\usepackage{lineno}
\usepackage{comment}
\usepackage{xspace}
\usepackage{dirtree}
\usepackage{enumitem}
\setlist[enumerate]{topsep=0pt,itemsep=0pt,partopsep=0pt,parsep=0pt}
\usepackage{bytefield}
\hypersetup{
    colorlinks=true,
    linkcolor=cyan,
    filecolor=magenta,      
    urlcolor=blue,
    citecolor=blue,
}

\newcommand{\PH}{\ensuremath{\mathrm{H}}\xspace}
\newcommand{\PQb}{\ensuremath{\mathrm{b}}\xspace}

\newcommand{\bbbar}{\ensuremath{\PQb\overline{\PQb}}\xspace}
\newcommand{\Hbb}{\ensuremath{\mathrm{H}\to\bbbar}\xspace}

\begin{document}

\title{FAIR AI Models in High Energy Physics}

\author{Javier Duarte$^{1}$, Haoyang Li$^{1}$, Avik Roy$^{2}$, Ruike Zhu$^{2,3}$, E.~A.~Huerta$^{3,4}$, Daniel Diaz$^{1}$, Philip~Harris$^{5}$, Raghav Kansal$^{1}$, Daniel~S.~Katz$^{2}$, Ishaan H. Kavoori$^{1}$, Volodymyr~V.~Kindratenko$^{2}$, Farouk Mokhtar$^{1,6}$, Mark~S.~Neubauer$^{2}$, Sang~Eon~Park$^{5}$, Melissa~Quinnan$^{1}$, Roger~Rusack$^{7}$, and Zhizhen~Zhao$^{2}$}

\address{$^{1}$University of California San Diego, La Jolla, California 92093, USA\\
$^{2}$University of Illinois at Urbana-Champaign, Urbana, Illinois 61801, USA\\
$^{3}$Argonne National Laboratory, Lemont, Illinois 60439, USA\\
$^{4}$The University of Chicago, Chicago, Illinois 60637, USA\\
$^{5}$Massachusetts Institute of Technology, Cambridge, Massachusetts 02139, USA\\
$^{6}$Hal{\i}c{\i}o\u{g}lu Data Science Institute, La Jolla, California 92093, USA\\
$^{7}$The University of Minnesota, Minneapolis, Minnesota 55405, USA}

\ead{jduarte@ucsd.edu}
\vspace{10pt}
\begin{indented}
    \item[]29 December 2023
\end{indented}

\begin{abstract}
    The findable, accessible, interoperable, and reusable (FAIR) data principles provide a framework for examining, evaluating, and improving how data is shared to facilitate scientific discovery.
    Generalizing these principles to research software and other digital products is an active area of research.
    Machine learning (ML) models---algorithms that have been trained on data without being explicitly programmed---and more generally, artificial intelligence (AI) models, are an important target for this because of the ever-increasing pace with which AI is transforming scientific domains, such as experimental high energy physics (HEP).
    In this paper, we propose a practical definition of FAIR principles for AI models in HEP and describe a template for the application of these principles.
    We demonstrate the template's use with an example AI model applied to HEP, in which a graph neural network is used to identify Higgs bosons decaying to two bottom quarks.
    We report on the robustness of this FAIR AI model, its portability across hardware architectures and software frameworks, and its interpretability.
\end{abstract}

\submitto{\MLST}
\maketitle

\section{Introduction}
\label{sec:intro}
Breakthroughs in machine learning (ML) and artificial intelligence (AI) have had a major impact on a range of scientific disciplines, including high energy physics (HEP), which is the study of the fundamental constituents of matter and their interactions.
In HEP, multiple experimental collaborations have used ML techniques extensively to address a broad range of problems.
For example, they were integral to the 2012 discovery of the Higgs boson~\cite{Chatrchyan:2012ufa,Aad:2012tfa} and subsequent observation of its decay to bottom quarks~\cite{CMS:2018nsn,ATLAS:2018kot} at the CERN Large Hadron Collider (LHC), where they were used to identify in proton-proton collisions the nature and origin of `jets' of particles produced in the collisions.
In another significant application, ML was used to identify in real time about 1000 events of interest from the forty million background events produced each second at the LHC~\cite{Duarte:2018ite,CMSP2L1T}. 
To maximize the scientific impact and utility of AI models in HEP, we propose a set of findable, accessible, interoperable, and reusable (FAIR) principles for them.

Our approach is inspired by community-wide initiatives that have produced guiding principles to maximize the reuse and scientific reach of digital assets.
Specifically, the FAIR principles were originally introduced~\cite{fairguiding} as guidelines for the management and stewardship of scientific datasets to optimize their reuse.
Recently, the FAIR4RS working group has developed an interpretation of the FAIR principles specifically for research software~\cite{FAIR4RS,KATZ2021100222,FAIR4RSFinal,FAIR_4_RS}, and FAIR principles have also been applied in the context of benchmarking and tool development~\cite{9652871}, and on the creation of computational frameworks for AI models~\cite{FAIR_diffraction}.

While these are important steps, these prior interpretations of FAIR principles are not readily applicable to AI models, which are conceptually and structurally different from data and research software.
Elucidating the details needed for a robust and general definition of FAIR principles for AI models requires application-specific benchmarks.
To address these challenges, we propose an operational definition of FAIR for HEP AI models, focusing on pre-trained models used to make predictions on HEP data.
These principles are intended to promote research reuse and reproducibility, which are known challenges in AI-driven scientific application research~\cite{transparency_reproducibility_ai}.
In addition, we present a method to automate the production, standardization, and publication of Python-based FAIR AI models in HEP.

To illustrate our proposed FAIR AI model definition in the context of HEP, we use a FAIR dataset to create and publish a FAIR AI model. 
Specifically, we use a simulated Higgs boson dataset distributed by the CMS Collaboration~\cite{ref:dataset,Chen:2021euv,McCauley:20197f}.
This FAIR dataset has been used for ML studies~\cite{Moreno:2019neq}, college courses~\cite{Benelli:2022sqn,UCSD-ParticalML} and tutorials~\cite{iaifi_summer_school_tutorials}.
We create a FAIR version of an interaction network (IN) AI model for 
Higgs boson identification~\cite{Moreno:2019neq}, and show how adopting our FAIR principles simplifies porting the model across different hardware architectures and software frameworks and facilitates the study of its interpretability.

This paper is organized as follows:
Section~\ref{sec:methods} outlines the methods used, where Section~\ref{sec:FAIRAI} describes related work and a formulation of FAIR principles for AI models; 
Section~\ref{sec:cookiecutter} introduces an AI project template; Section~\ref{sec:mapping} summarizes how the template maps to FAIR principles, and 
Section~\ref{sec:implementation} describes an example of the application of FAIR principles, where we take a previously published AI model in HEP~\cite{Moreno:2019neq} and make it FAIR.
Next, Section~\ref{sec:results} discusses the portability and interpretability of this model, as enabled by the FAIR principles.
Finally, Section~\ref{sec:summary} summarizes the paper.

\section{Methods}
\label{sec:methods}
\subsection{FAIR principles for AI models in HEP}
\label{sec:FAIRAI}
Substantial work has been done to investigate how to apply
the FAIR principles to research software~\cite{FAIR4RS,KATZ2021100222,FAIR4RSFinal,FAIR_4_RS}.
The design, optimization, and training of ML models combine disparate digital assets, including research software, data, libraries and tools, workflows, and an expanding ecosystem of hardware architectures.
Depending on the use case, AI models can often be optimized to be faster, more parallel, or better utilize the underlying hardware within different software toolkits.
To minimize misinterpretation, the reproducibility and reusability of AI models require details of provenance for the entire discovery cycle.
In addition, to execute the AI model on a new dataset, including new data that has not been preprocessed, an exact recipe of the data preparation and preprocessing steps is required, such as the units used to express the data features~\cite{units}.

Operationally, an AI model is usually instantiated in a software framework, such as Scikit-learn~\cite{scikit-learn}, TensorFlow~\cite{tensorflow}, PyTorch~\cite{pytorch}, XGBoost~\cite{xgboost}, or ONNX~\cite{onnx}, that may be serialized in a file on disk.
The storage of models within these formats can vary from low-level hardware-optimized intermediate representations (IRs) to high-level IRs, leading to different inference results or performance.
In addition, preparation and prepossessing steps, which can have an impact on the model, can be specified either in separate scripts, or as layers integrated into the model.
There are efforts to share such code as open-source GitHub repositories, like Papers with Code~\cite{paperswithcode}.
However, it has been observed that these repositories are often incomplete, lacking key information, and not maintained, making the results difficult to reproduce~\cite{WATTANAKRIENGKRAI2022111117,JMLR:v22:20-303,transparency_reproducibility_ai}.
This has led to the establishment of AI reproducibility challenges~\cite{JMLR:v22:20-303,rc2022}.
In light of these considerations, we propose the following definition for a FAIR AI model, aimed at meeting the high-level goals of F, A, I, and R (the four foundational principles) in the original FAIR data principles~\cite{fairguiding} for AI models:
\begin{center}
    \fbox{\begin{minipage}{0.9\textwidth}
            \textbf{\emph{An AI model consists of the architecture (computational graph) and a given set of parameters, which can be expressed as source code files or executables needed to run inference (i.e., produce outputs) on a data sample.
                    A FAIR AI model is an AI model that satisfies the properties listed in Table~\ref{tab:FAIR-defs}.
                    In brief, (F) the model and its associated metadata are easy to find for both humans and machines, (A) the model and its metadata are retreivable via standardized protocols, (I) the model interoperates with other models, data, and/or software, and (R) the model is both usable and reusable.}}
        \end{minipage}}
\end{center}

For an ML model to be FAIR, we stress that, first, the dataset used to train the model must be FAIR, and follow domain-relevant community standards, because the dataset is an essential part of the ML model's provenance.
In \tablename~\ref{tab:FAIR-defs}, we present a set of proposed FAIR AI principles, adapted from the FAIR principles created for research software~\cite{FAIR4RSFinal} by the Research Data Alliance (RDA) FAIR for Research Software (FAIR4RS) Working Group~\cite{FAIR4RS,KATZ2021100222,FAIR4RSFinal,rda,katz_daniel_s_2022_6647819}.
This set of principles has been given to the RDA FAIR for ML (FAIR4ML) interest group~\cite{RDA-FAIR4ML} that formed in September 2022.
We believe that these guidelines are the minimum criteria for a model to be considered as FAIR.
However, additional criteria may be necessary to truly ensure a shareable, reproducible, and extendable ML model.

A critical challenge to ensure reproducibility is that of backend optimizations.
The output of the AI algorithm can be affected by changes in the operation order, operation precision, and parallelization strategy.
Currently, frameworks such as PyTorch and ONNX have different intermediate representations (IRs), which can lead to different outputs depending on how the model is initialized or compiled.
These differences can be substantial even when the same hardware is used~\cite{pytorch_github_issue}.
Moreover, specific processor types may have limitations in the bit precision of various operations.
Differences in precision can lead to substantial deviations, rendering exact reproducibility across processors nearly impossible.
As a consequence, for the purposes of this discussion, we refer to reproducibility as the ability to produce results that are statistically consistent with the aggregate data on a large scale, but when comparing a single inference on the same data, can deviate within a specified tolerance.

\begin{table}[htbp]
    \small
    \caption{Proposed FAIR principles for fully trained AI models used for AI-inference only, based on adapting the original FAIR principles by initially replacing data by AI models and then making further changes based on the characteristics of AI models versus datasets and the ways they are developed, shared, searched for, and used.
        These proposed principles could be further extended for retraining use cases by amending our proposed definition for the `Reusability' principle.}
    \label{tab:FAIR-defs}
    \centering
    \begin{tabular}{|p{0.9\textwidth}|}
        \hline
        \textbf{F: The AI model, and its associated metadata, are easy to find for both humans and machines.}                                                                                                                      \\\hline
        \begin{enumerate}
            \item[F1.] The AI model is assigned a globally unique and persistent identifier.
            \item[F2.] The AI model is described with rich metadata.
            \item[F3.] Metadata clearly and explicitly include the identifier of the AI model they describe.
            \item[F4.] Metadata and the AI model are registered or indexed in a searchable resource.
        \end{enumerate}                                                                                                                                                                                                  \\\hline
        \textbf{A: The AI model, and its metadata, are retrievable via standardized protocols.}                                                                                                                                    \\\hline
        \begin{enumerate}
            \item[A1.] The AI model is retrievable by its identifier using a standardized communications protocol.
                \begin{enumerate}
                    \item[A1.1.] The protocol is open, free, and universally implementable.
                    \item[A1.2.] The protocol allows for an authentication and authorization procedure, where necessary.
                \end{enumerate}
            \item [A2.] Metadata are accessible, even when the AI model is no longer available.
        \end{enumerate}                                                                                                                                                                                                  \\\hline
        \textbf{I: The AI model interoperates with other models, data, and/or software by exchanging data and/or metadata, and/or through interaction via application programming interfaces (APIs), described through standards.} \\\hline
        \begin{enumerate}
            \item[I1.] The AI model reads, writes and exchanges data in a way that meets domain-relevant community standards.
            \item[I2.] The AI model includes qualified references to other objects, including the (FAIR) data used to train the model.
        \end{enumerate}                                                                                                                                                                                                  \\\hline
        \textbf{R: The AI model is both usable (for inference) and reusable (can be understood, built upon, or incorporated into other models and/or software).}                                                                   \\\hline
        \begin{enumerate}
            \item[R1.] The AI model is described with a plurality of accurate and relevant attributes.
                \begin{enumerate}
                    \item[R1.1.] The AI model is given a clear and accessible license.
                    \item[R1.2.] The AI model is associated with detailed provenance, such as information about the input data preparation and training process.
                \end{enumerate}
            \item[R2.] The AI model includes qualified references to other models and/or software, such as dependencies.
            \item[R3.] The AI model meets domain-relevant community standards.
        \end{enumerate}                                                                                                                                                                                                  \\\hline
    \end{tabular}
\end{table}

\subsection{Cookiecutter4fair: FAIR AI project template}
\label{sec:cookiecutter}
Software templates can be used to encourage good practices; Cookiecutter Data Science~\cite{cookiecutter_data_science} is one such template that is specifically oriented at data science projects.
It consists of a logical, reasonably standardized, but flexible project structure hosted on GitHub for performing and sharing data science work.
We took inspiration from this and created a fork of this template generator, called \texttt{cookiecutter4fair}~\cite{cookiecutter4fair}, with additional features to promote the adoption of our FAIR principles.
Other tools, like Showyourwork~\cite{show_your_work}, specifically address the issue of reproducibility in science.

\subsubsection{Usage}

\begin{figure}[htbp]
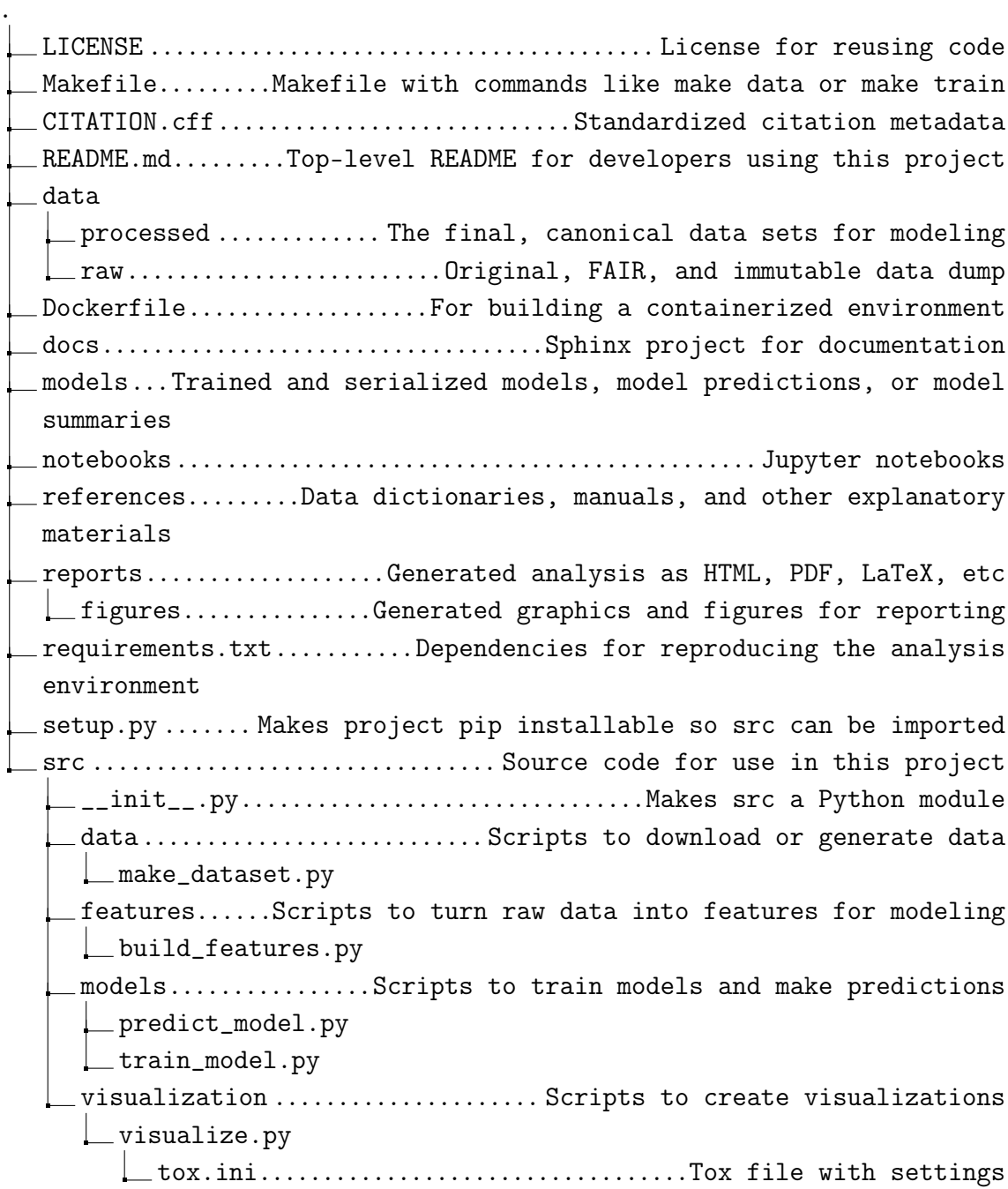

    \dirtree{%
        .1 ..
        .2 LICENSE\dotfill License for reusing code.
        .2 Makefile\dotfill  Makefile with commands like make data or make train.
        .2 CITATION.cff\dotfill Standardized citation metadata.
        .2 README.md\dotfill Top-level README for developers using this project.
        .2 data.
        .3 processed\dotfill The final, canonical data sets for modeling.
        .3 raw\dotfill Original, FAIR, and immutable data dump.
        .2 Dockerfile\dotfill For building a containerized environment.
        .2 docs\dotfill Sphinx project for documentation.
        .2 models\dotfill  Trained and serialized models, model predictions, or model summaries.
        .2 notebooks\dotfill Jupyter notebooks.
        .2 references\dotfill Data dictionaries, manuals, and other explanatory materials.
        .2 reports\dotfill Generated analysis as HTML, PDF, LaTeX, etc.
        .3 figures\dotfill Generated graphics and figures for reporting.
        .2 requirements.txt\dotfill Dependencies for reproducing the analysis environment.
        .2 setup.py\dotfill Makes project pip installable so src can be imported.
        .2 src\dotfill Source code for use in this project.
        .3 \_\_init\_\_.py\dotfill Makes src a Python module.
        .3 data\dotfill Scripts to download or generate data.
        .4 make\_dataset.py.
        .3 features\dotfill Scripts to turn raw data into features for modeling.
        .4 build\_features.py.
        .3 models\dotfill Scripts to train models and make predictions.
        .4 predict\_model.py.
        .4 train\_model.py.
        .3 visualization\dotfill Scripts to create visualizations.
        .4 visualize.py.
        .5 tox.ini\dotfill Tox file with settings.
    }
    \caption{
        Folder hierarchy of the \texttt{cookiecutter4fair}~v1.0.0~\cite{cookiecutter4fair} project template.
        The main Python source code is contained in \texttt{src}.
        The \texttt{docs} folder contains a Sphinx project for generating documentation.
    }
    \label{fig:template}
\end{figure}

The project template is designed to be used with the \texttt{cookiecutter}~\cite{cookiecutter} program, a command-line utility that creates projects from project templates using the Jinja2~\cite{jinja} templating engine, and that can be installed via \texttt{pip}.
A new FAIR AI project can be made with the command \texttt{cookiecutter https://github.com/FAIR4HEP/cookiecutter4fair}.
The first argument corresponds to the project template that is hosted on GitHub.
After asking the user for the project name, repository name, author name, author ORCID, description of the project, chosen license, DOI for the input data, DOI for the code (if available), and whether to include a template Dockerfile, \texttt{cookiecutter} will create the template structure as shown in Fig.~\ref{fig:template}.

The questions that the repository asks the user upon project creation can be found and modified in the file \texttt{cookiecutter.json}.
The \texttt{Makefile} contains commands that allow the user to do various things with their project, such as downloading the data, setting up the test environment, converting the dataset, and training and evaluating the model.
It also contains global variables obtained from \texttt{cookiecutter.json}.
This procedure makes it explicit that the analysis operations are a directed acyclic graph (DAG).

If the data is hosted on Zenodo~\cite{zenodo}, the user can download the data from the DOI link by invoking \texttt{make sync\_data\_zenodo}, which uses the \texttt{zenodo\_get} command line utility~\cite{david_volgyes_2020_10.5281/zenodo.1261812} to download the data.
The \texttt{Dockerfile} can be built and run to provide a Python environment for the project to work, which installs the dependencies specified in \texttt{requirements.txt}.
When the Docker image is built, it can be run interactively with the command \texttt{docker run -d -t <image name>}.
The pre-project and post-project scripts are automatically run before and after the project directory is generated and provide additional flexibility.
After the project template has been generated, the user can organize their source code and documentation in order to follow the FAIR principles.

\subsubsection{Design considerations based on FAIR principles}

\paragraph{Findable}

There are many ways to ensure findability for AI models once they are created and published.
Simple ways include uploading it to GitHub, GitLab, or BitBucket.
Several efforts aim to create ``model commons,'' hubs in which models can be shared.
Among these are DLHub~\cite{DLHub,chard2016globus}, OpenML~\cite{openml}, MLCommons~\cite{mlcommons}, AI Model Share~\cite{modelshare}, and Hugging Face~\cite{Wolf_Transformers_State-of-the-Art_Natural_2020}.
If a publication or arXiv preprint is associated with the software, the code repository can also be linked to it via Papers with Code~\cite{paperswithcode}.
However, this does not really support the findability principle.

To improve findability, Zenodo~\cite{zenodo} can be leveraged to generate a DOI for the repository, as well as to store metadata.
Recently, HuggingFace also enabled the ability to generate DOIs for both data sets and models~\cite{huggingface_doi}.
Ideally, we would like a way to search all these repositories at the same time.
This would require that they each expose a machine accessible search mechanism, ideally using a common standard, and that there is a way to perform a federated search across the full set of repositories.

\paragraph{Accessible}
Accessibility is another place where standardization is needed.
Specifically, we need a standard, open, free, protocol for retrieving a model from an identifier.
Then the various model repositories would need to implement the server side of this protocol, and community members would likely then implement the client side of the protocol in common tools in Python, R, and other programming languages.

\paragraph{Interoperable}
To ensure interoperability, the metadata describing the AI model must thoroughly document all aspects of its structure, training, and inputs, including any prepossessing needed for the raw data and a provenance of the data.
To enable machine interoperability, standardized APIs, such as those associated with DLHub, HuggingFace, or NVIDIA Triton Server, can be used~\cite{triton}.

\paragraph{Reusable}
To enable reusability, it is important to specify the software, tools, and dependencies needed to seamlessly invoke an AI model to extract knowledge from datasets in a given computing environment.
This process should be hardware agnostic.
This may be accomplished by using container solutions, such as Docker~\cite{docker} or Apptainer~\cite{singularity_osti}.

Reusability for inference only requires fully trained ML models.
In this context, a trained ML model may be reusable as the backbone to develop another model or to fine-tune it to perform a different task, e.g., the WaveNet model~\cite{2016wavenet}, originally developed for text-to-speech and music generation has been adapted for classification and regression tasks in  astrophysics~\cite{2021NatAs.tmp..118H,2022PhLB..83537505K}.
Recent approaches based on ``foundation models,''~\cite{foundation_models} in which large models (sometimes containing up to $10^9$ parameters) are pre-trained on unlabeled datasets and subsequently fine-tuned for downstream tasks, illustrate the need for reusability at large scale.
These approaches envision the creation of a small collection of general-purpose AI models that may be reused for a large class of tasks.

\paragraph{Other considerations}
Optimally deploying models on a given hardware processor often involves modifying the internal structure of the model to better utilize the hardware resources.
These optimizations correspond to transformations of IRs, specified, e.g., in ONNX or the more flexible Multi-Level IR (MLIR)~\cite{MLIR}.
These transformations can change the numerical output values of models, affecting their reproducibility.
There has been limited broad scale acceptance of a standard IR for AI models.
In place of this, appropriate metadata describing the hardware used and any hardware-specific optimizations is needed to ensure the model can be reliably reproduced.

In some ways, a higher standard than FAIR is full reproducibility.
To ensure reproducibility requires clearly communicating the details of the full end-to-end AI cycle encompassing data collection and curation, API selection for model R\&D, hyperparameter optimization, design of domain-inspired loss functions, distributed training schemes, optimizers, random/frozen initialization of weights, data split choices for training, validation, testing and quantization, data loaders, hardware used, and hardware-specific optimizations, among other details.
The diverse and rather disparate portfolio of available choices, and the different levels of AI  and computing skills of end users, may mean that full reproducibility is not possible.
In this article, we propose a minimum and achievable standard of FAIR principles in the context of AI models used for inference.

\subsection{Mapping to FAIR principles}\label{sec:mapping}
Table~\ref{tab:FAIR} summarizes how the features of the \texttt{coookiecutter4fair} AI project template map to the proposed FAIR principles for AI models.
Most aspects are fully automated, such as the creation of a license file and \texttt{Dockerfile} for creating an environment.
Some aspects are partially automated, such as uploading the model to Zenodo.
In particular, the GitHub--Zenodo bridge can be enabled from the Zenodo web interface, which automates the generation of an updated entry for each new release on GitHub.
The \texttt{coookiecutter4fair} repository template populates a \texttt{CITATION.cff} file~\cite{Druskat_Citation_File_Format_2021} with citation metadata, which can then be used by Zenodo.
Finally, other aspects are not fully automated, but require some additional manual steps, such as uploading the model to DLHub as described above.

\begin{table}[!htpb]
    \small
    \centering
    \caption{
        Map between existing capabilities of the \texttt{coookiecutter4fair} AI project template and our proposed FAIR principles for AI models.
        The $\ast$ symbol indicates that the process is not yet fully automated and requires additional manual steps.
    }
    \begin{tabular}{p{2.2cm}|p{2cm}p{2cm}p{2cm}p{3cm}p{1.8cm}}
        Principle     & GitHub\newline repository & Zenodo\newline upload & DLHub\newline upload & Docker or\newline Apptainer image & License      \\
        \hline
        Findable      & $\checkmark$              &                       &                      &                                   &              \\
        Accessible    &                           & $\checkmark$          & $\ast$               &                                   &              \\
        Interoperable &                           &                       &                      & $\checkmark$                      &              \\
        Reusable      &                           &                       & $\ast$               & $\checkmark$                      & $\checkmark$
        \\
    \end{tabular}
    \label{tab:FAIR}
\end{table}

\subsection{\texorpdfstring{FAIR implementation of \Hbb interaction network}{FAIR implementation of H(bb) interaction network}}
\label{sec:implementation}
The Higgs boson is a linchpin of the standard model (SM) of particle physics.
It is a byproduct of the mechanism that generates masses for all elementary particles.
Studying its properties, such as its production and decay rates, is one of the overarching goals of the CERN LHC program, and any deviations measured with respect to the SM may give a hint to elusive new physics.
The Higgs boson most commonly decays (about 58\% of the time) to a bottom quark-antiquark pair ($\bbbar$).
Traditionally, this is a difficult decay of the Higgs boson to study because there is a large background consisting of jets produced through the strong interactions.
These are known as quantum chromodynamics (QCD) multijet events.
ML models, especially graph neural networks (GNNs)~\cite{Qu:2019gqs,Moreno:2019neq}, have been shown to dramatically improve the rejection of this background, while retaining high $\PH\to\bbbar$ detection efficiency thus enabling the study of this decay mode.
In this section, we provide a concrete example of implementing one such model, which is an interaction network (IN) model described in Ref.~\cite{Moreno:2019neq}, following our recommendations for a FAIR AI model.

\begin{figure}[ht]
  \centering
  \includegraphics[width=0.7\textwidth,viewport=400 200 1520 880,clip=true]{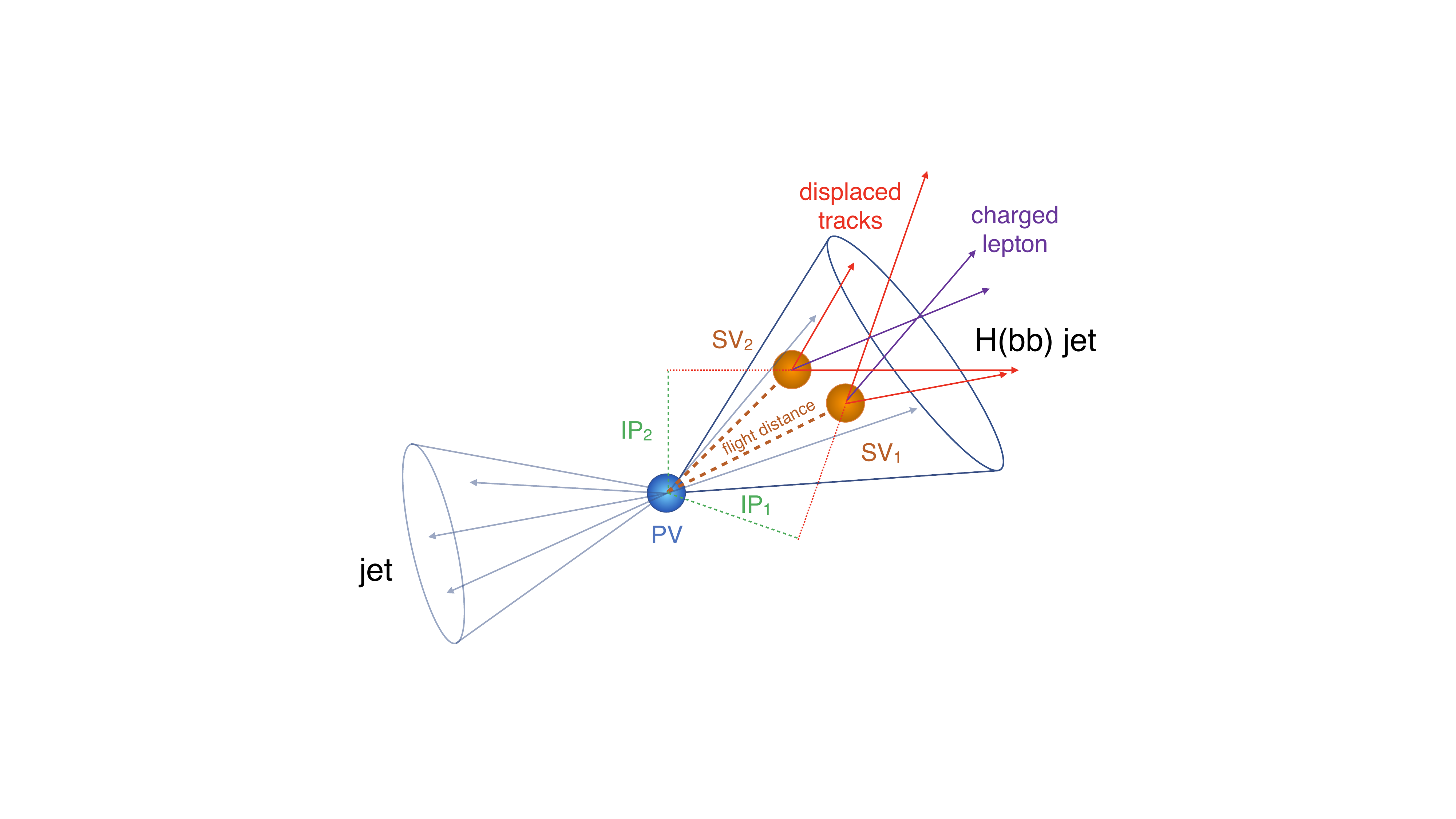}
  \caption{Illustration of a $\Hbb$ jet with two secondary vertices (SVs) from the decay of two bottom hadrons resulting in charged-particle tracks (including a low-energy, or soft, lepton) that are displaced with respect to the primary collision vertex (PV), and hence have a large impact parameter (IP) value.}
  \label{fig:HbbTagging}
\end{figure}

The data structure in HEP is defined around the concepts of events.
These are discrete moments where all the particles arising from a single proton-proton collision are measured by a detector and recorded.
Each event is independent of all the other events.
A dataset may consist of several millions of events.
To identify events with a $\Hbb$ decay and separate them from the much larger QCD background, several salient features are illustrated in Fig.~\ref{fig:HbbTagging}.
At the LHC, for each event particle candidates are reconstructed from detector measurements and clustered into cone-shaped jets, attempt capture most of the energy from a single particle produced in the collision, such as Higgs boson.
Charged particles produced in the collision are detected and the momenta and direction are measured in a tracking detector.
These tracks are collected to form jets.
There is a special class of jets from bottom quarks where the particles travel a measurable distance from the collision vertex before decaying to other particles, forming a so-called secondary vertex (SV).
It is this class of jets that we are searching for when we search for $\PH\to\bbbar$ decays.

\subsubsection{Interaction network model}

\begin{figure}[htbp]
  \centering
  \includegraphics[width=1\textwidth]{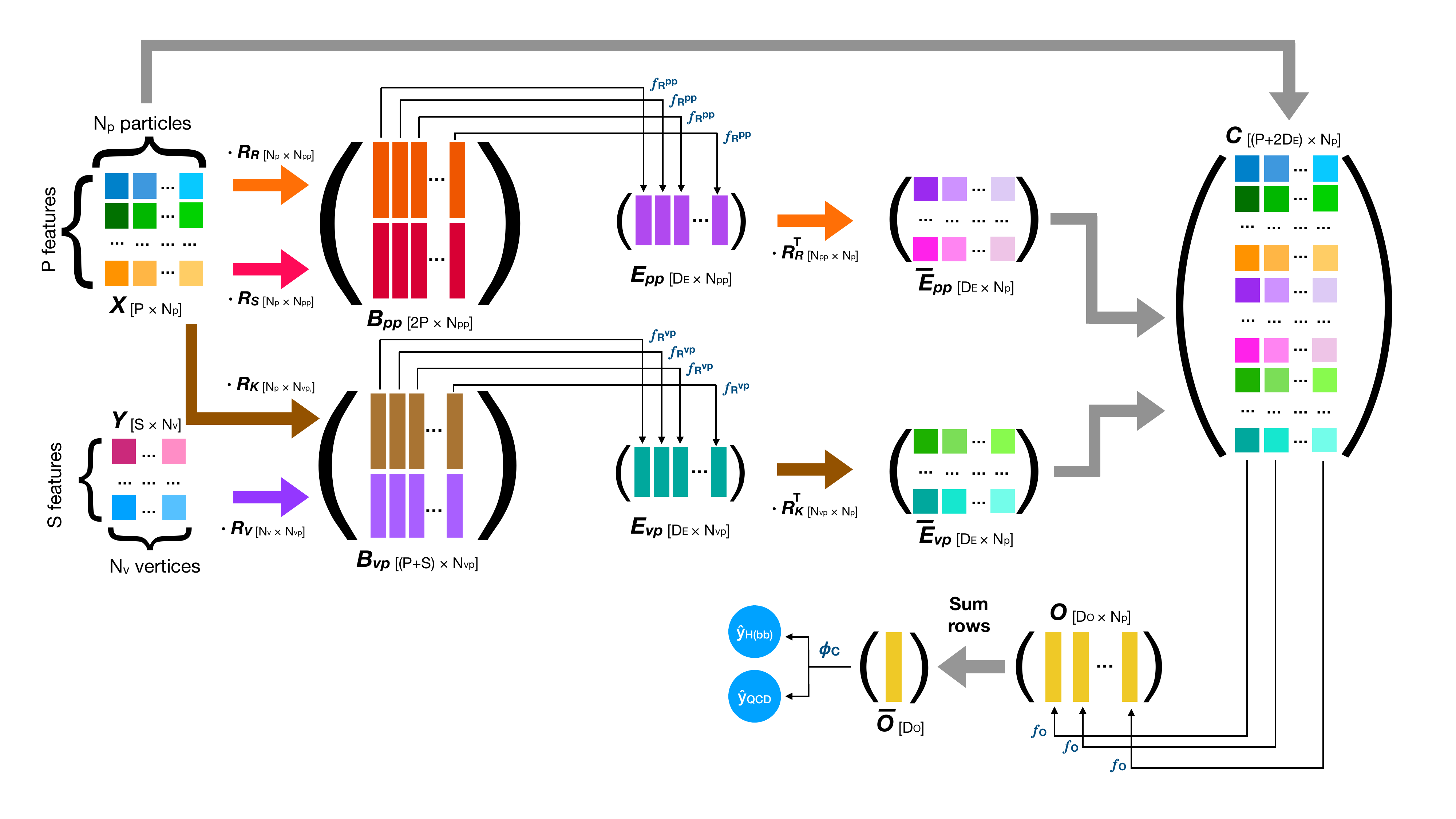}
  \caption{
    Network architecture and dataflow in the IN model~\cite{Moreno:2019neq}.
    The choice of model hyperparameters and input data dimensions for the baseline model is given in the accompanying table.
  }
  \label{fig:IN-arch}
\end{figure}

The IN model was first proposed~\cite{Battaglia:2016jem} in order to explore the evolution of physical dynamics and was later adapted for the task of jet classification; in this case differentiating \Hbb jets from QCD jets~\cite{Moreno:2019neq}.
The dataset for training, validation, and testing is derived from the CMS open simulated dataset with 2016 conditions that is available from the CERN Open Data Portal~\cite{ref:dataset}.
It consists of jets, decomposed into constituent charged particle tracks, and SVs, labeled as either \Hbb signal or QCD background.
More information on the dataset can be found in Chen et al.~\cite{Chen:2021euv}.
Figure~\ref{fig:IN-arch} shows the IN model architecture and Table~\ref{tab:IN-hyperparameters} provides the values of the model hyperparameters as well as input data dimensions for the baseline model.
For a detailed description of the model and chosen hyperparameters, see Moreno et al.~\cite{Moreno:2019neq}

As discussed in Moreno et al.~\cite{Moreno:2019neq}, graphs are natural data structures to describe jets because they are permutation invariant (i.e., there is no preferred order to the constituents of the jet), they can accommodate variable-sized objects (i.e., jets may be composed of a few or many constituents), and they can describe entities as nodes (i.e., constituents) and their relations as edges.
This network was trained on graph data structures based on up to $N_p=30$ particle tracks, each with $P=60$ features, and up to $N_v=5$ SVs, each with $S=14$ features, associated with the jet.
The physical description of each feature is given in Appendix C of Moreno et al.~\cite{Moreno:2019neq}.

Two input graphs are used: a fully-connected directed graph with $N_{pp} = N_p(N_p - 1)$ edges between the particle tracks and a separate graph with $N_{vp} = N_vN_p$ connections between the particle tracks and the SVs.
The node level feature space of the fully connected track graph is transformed to edge level features via two interaction matrices, identified as $R_{R[N_p \times N_{pp}]}$ and $R_{S[N_p \times N_{pp}]}$, where the former accounts for how each node receives information from other nodes and the latter encodes the information about each node sending information to other nodes.
The track-vertex graph is transformed by similarly defined interaction matrices: $R_{K[N_p \times N_{vp}]}$ and $R_{V[N_v \times N_{vp}]}$.
The feature spaces of these graphs are transformed via nonlinear functions, respectively called $f_R^{pp}$ and $f_R^{vp}$, to obtain two $D_E$ dimensional internal state representations of these graphs.
These nonlinear functions are approximated by fully connected multilayer perceptrons (MLPs).

These internal state representations, respectively given by $E_{pp[D_E \times N_{pp}]}$ and ${E}_{vp[D_E \times N_{vp}]}$ matrices, are transferred back to the particle tracks by transforming them with $R_R^T$ and $R_K^T$ matrices.
These transformed particle level representations are given by matrices $\bar{E}_{pp[D_E \times N_{p}]}$ and $\bar{E}_{vp[D_E \times N_{p}]}$ respectively.
Concatenating these particle-level internal state representations with the original track features creates a feature space with a dimension of $(P+2D_E)$ for each of the  $N_p$ tracks.
The function $f_O$, represented by a trainable dense MLP, creates the post-interaction $D_O$ dimensional internal representation that is stored in the matrix $O_{[D_O \times N_p]}$.
Finally, these track-level internal representations are summed to obtain a $D_O$ dimensional state vector $\bar{O}$ and linearly combined to produce a two-dimensional output, which is transformed to individual class probabilities via a softmax function.

\begin{table}[htbp]
  \centering
  \caption{The choice of IN model hyperparameters and input data dimensions for the baseline model.}
  \begin{tabular}[b]{ l|c }
    Hyperparameter         & Value             \\
    \hline
    $(P, N_p, S, N_v)$     & $(30, 60, 14, 5)$ \\
    No. of hidden layers   & 3                 \\
    Hidden layer dimension & 60                \\
    $(D_E, D_O)$           & (20, 24)          \\
    Activation             & ReLU              \\
  \end{tabular}
  \label{tab:IN-hyperparameters}
\end{table}

\subsubsection{FAIR implementation}

We created a FAIR implementation of the AI model hosted on GitHub and Zenodo~\cite{hbb_interaction_network}.
The repository was initialized using the template described in Section~\ref{sec:cookiecutter}.

\paragraph{Features}

The repository includes a dataset processing script that converts the raw data from the CERN Open Data portal.
It also has training and prediction scripts to reproduce the published results.
As described above, \texttt{Makefile} contains all of these commands, which codifies the analysis as a DAG.

In addition, two \texttt{Dockerfiles} that can create reproducible environment for either CPU-based or GPU-based model training and inference are included in the repository.
These images are prebuilt and hosted on DockerHub.
We also automated documentation generation, training and inference workflows, Docker container building, with continuous integration through GitHub Actions.
Finally, a DOI is generated using the Zenodo--GitHub bridge, in which a new DOI is minted for each new release of the software on GitHub.

\paragraph{Deployment to DLHub}

We have made the trained ML model accessible~\cite{dlhubmodelin} and reusable for inference by making it publicly available via DLHub~\cite{chard2019dlhub,DLHub}.
DLHub provides a custom software development kit (SDK) called \texttt{dlhub\_sdk} that allows users to package and preserve a trained model with necessary dependencies, including packages with specific versions, custom modules, and serialized data and model files.
Once a model has been published, its dedicated API can be used to run remote inference tasks using \texttt{funcX}, a fire-and-forget remote function execution that elastically deploys workers and containers across nodes in clouds, clusters, and supercomputers~\cite{chard2020funcx}.
The process of making a model available is simplified with a notebook template made available by DLHub developers.
This notebook requires the user to implement the inference code as a function that is executed during model calls, and to declare model-specific dependencies and associate metadata.
The notebook template is accompanied with a document template with necessary information about the model.
The prescription of using these templates is user friendly: once both templates are filled out and the notebook successfully runs, they can be sent to the DLHub developers who streamline the process of depositing and curating the model.
The published model includes a DOI, list of authors, point of contact, relevant information about input and output data type and shape, and instructions to run the ML model with a sample test set.
DLHub's SDK also allows users to explore the model's metadata, which encompasses dependencies and libraries used to create and containerize the model, and information about the tasks performed by the model, e.g., classification or regression.

\section{Results}
\label{sec:results}
\subsection{Portability and performance across platforms}
\label{sec:portability}
In this section, we examine the portability and extensibility of the IN model, a graph neural network used for the jets classification task.
In Section~\ref{sec:reproducibility}, we reproduce the training and evaluation of the IN model with the same hyperparameters and dataset as Moreno et al.~\cite{Moreno:2019neq}.
Section~\ref{sec:robustness} retrains the model with different training-validating splits on different servers to test the reproducibility of the results under different conditions.
In Sections~\ref{sec:portability_hardware} and~\ref{sec:portability_software}, we explore the model's portability across software frameworks and hardware platforms.
We convert the model from PyTorch to TensorRT, using ONNX as the intermediate format, and evaluate the model's inference speed and compatibility of results.
We also create an Apptainer container~\cite{singularity_osti} to improve the model's portability across platforms, and evaluate the model's inference performance within the container.

\subsubsection{Reproducibility}
\label{sec:reproducibility}

In this subsection we provide details of training the benchmark experiments of the IN model with the same data input and hyperparameters setting as used by Moreno et al.~\cite{Moreno:2019neq}.
The training samples are saved in 57 HDF5 files, each of which contains about 100\,k jets.
We use 52 of them for training and 5 for validation.
The testing dataset is saved as a set of NumPy array files (one feature per file), where each file contains 600\,k jets.

There are several differences in our experiment setting compared to Moreno et al.
For the training platform, we use the Hardware Accelerated Learning (HAL) GPU cluster at the National Center for Supercomputing Applications
(NCSA)~\cite{10.1145/3311790.3396649} as a remote GPU cluster and train on the NVIDIA V100 GPU, while Moreno et al. trained their model on one NVIDIA GeForce GTX 1080 GPU.
For the data splitting, we take the first five HDF5 files as validation data and the rest as training data.
Moreno et al. split the data into training, validation, and test samples, with 80, 10, and 10\% of the data respectively.
In our training process, each epoch takes about 450\,s to finish.
The training terminates following the early stopping condition when the validation loss failed to improve for 8 epochs.
As a first check, Table~\ref{tab:benchmark} shows a comparison of our training results and the results from Moreno et al.
We repeat the training 10 times varying the random seed used for initialization and data shuffling, and report the mean and standard deviation of the validation accuracy and the AUC.
We also report the one-sided (upper tail) $p$-value for the original model given the distribution of our trials.
We find the reported performance of the original model is consistent ($p$-value $>5\%$) with our reproduction.

\begin{table}[htpb]
  \centering
  \caption{
    The IN model's performance in this work and as reported in the original publication.
    In this work, we repeat the training 10 times varying the random seed used for initialization and data shuffling, and report the mean and standard deviation of the validation accuracy and the AUC.
    We also report the one-sided (upper tail) $p$-value for the original model given the distribution of our trials.
    We find the reported performance of the original model is consistent ($p$-value $>5\%$) with our reproduction.
  }
  \begin{tabular}{l|lll}
                                             & Validation accuracy & AUC               \\
    \hline

    IN: This work                            & $0.9545\pm0.0005$   & $0.9898\pm0.0002$ \\
    IN: Original model~\cite{Moreno:2019neq} & 0.9550              & 0.9900            \\
    \hline
    $p$-value (consistency)                  & 12.69\%             & 9.27\%            \\
  \end{tabular}
  \label{tab:benchmark}
\end{table}

\subsection{Robustness}
\label{sec:robustness}

There are a variety of methods to quantify the stability of AI models.
Smart data samplers may be developed to expose ML models to novel information at every training epoch.
This may be a particularly challenging task if the parameter space is largely unknown, and the optimizer, loss function, and architecture do not encode domain information to properly constrain the ML model during the training stage.
Even if the method used to sample the parameter space under consideration during the training stage is suboptimal, the ML model may eventually converge and attain optimal performance, even if the training stage takes longer.
The performance of the fully trained model, however, should not be uniquely determined by the method used to split the training, validation, and test sets.
In fact, an optimal model should be robust to the selection of training, validation, and test sets, unless the information contained in these datasets is not representative of the phenomena that it aims to describe.

In view of these considerations, we have explored three different data split approaches to handle the HDF5 files that contain the jet data used to produce a new version of the IN model in this article, namely:
\begin{enumerate}
  \item Use $k$-fold cross-validation at the file level.
        In this approach, the data are split into folds, each containing five files.
        For training purposes, we select a $k$-fold as validation data and the rest
        as training data.
        We iterate over the entire dataset, and then calculate the average score of all training rounds.
  \item Randomly select five files as the validation set and the rest as the training set.
  \item Save the entire dataset as one NumPy array on disk and use the split function in Scikit-learn to randomly split the dataset to create training and validation sets.
\end{enumerate}

We explored these approaches using the IN model in the HAL cluster.
Our findings are summarized in Table~\ref{tab:different_split}.
Briefly, the IN model is robust to any of the different methods used to train it, which furnishes evidence for its stability and reliability.

\begin{table}[htbp]
  \centering
  \caption{
    Stability of IN model against different training methods in the HAL GPU cluster.
  }
  \resizebox{\textwidth}{!}{
    \begin{tabular}{|c|c|c|c|c|c|}
      \hline
      Training epochs & Valid. accuracy & Valid. loss & Train. (valid.) time/epoch [s] & AUC    \\
      \hline
      \multicolumn{5}{|c|}{Method 1: Cross validation}                                          \\
      \hline
      55              & 0.9541          & 0.1207      & 468.8 (23.0)                   & 0.9897 \\
      \hline
      \multicolumn{5}{|c|}{Method 2: Random split on file names}                                \\
      \hline
      71              & 0.9555          & 0.11757     & 445.4 (28.6)                   & 0.9901 \\
      \hline

      \multicolumn{5}{|c|}{Method 3: Random split on data points}                               \\
      \hline
      71              & 0.95546         & 0.11760     & 420.4 (20.5)                   & 0.9901 \\
      \hline
    \end{tabular}
  }
  \label{tab:different_split}
\end{table}

\subsection{Portability across hardware platforms}
\label{sec:portability_hardware}

To demonstrate the portability of our IN model implementation across different hardware architectures, we used the HAL and DGX systems at NCSA and the ThetaGPU supercomputer at the Argonne Leadership Computing Facility.
The specifications of each of these platforms are summarized in Table~\ref{tab:platform_table}.

\begin{table}[htbp]
  \centering
  \caption{Specifications of the DGX, HAL, and ThetaGPU systems.}
  \resizebox{\textwidth}{!}{
    \begin{tabular}{|c|c|c|c|}
      \hline
                       & HAL System               & DGX system         & ThetaGPU system    \\
      \hline
      Operation system & ppc64le RHEL 8.4 Linux
                       & x86\_64 CentOS 7.4 Linux & x86\_64 GNU/Linux                       \\
      \hline
      Package versions
                       & Python 3.7.10            & Python 3.6.8       & Python 3.8.10      \\
                       & PyTorch 1.7.1            & PyTorch 1.9.1      & PyTorch 1.10.0     \\
                       & Torchvision 0.8.2        & Torchvision 0.10.1 & Torchvision 0.11.1 \\
                       & Tqdm  4.59.0             &                    &                    \\
                       & Scikit-learn 0.24.2      &                    &                    \\
                       & h5py 3.2.1               &                    &                    \\
      \hline
      GPU              & NVIDIA V100              & NVIDIA A100        & NVIDIA A100        \\
      \hline
    \end{tabular}
  }
  \label{tab:platform_table}
\end{table}

Our IN model implementation is produced using a CMS dataset with a suitable format to fit the model's input data size and type.
Each file in the dataset includes $10^5$ data points.
\tablename~\ref{tab:platform_table} provides results for each of the three training methods described in the previous section, in each of the three high performance computing platforms used for this exercise.
Our findings indicate that our IN model implementation is hardware agnostic.

\begin{table}[htbp]
  \centering
  \caption{
    IN model portability is showcased using three different data split methods across three different high performance computing platforms.
  }
  \resizebox{\textwidth}{!}{
    \begin{tabular}{|c|c|c|c|c|c|c|}
      \hline
      Platform & Training epochs & Valid. accuracy & Valid. loss & Train. (valid.) time/epoch [s] & AUC    \\
      \hline
      \multicolumn{6}{|c|}{Method 1: Cross validation}                                                     \\
      \hline
      HAL      & 55              & 0.9541          & 0.1207      & 468.8 (23.0)                   & 0.9897 \\
      DGX      & 86              & 0.9563          & 0.1160      & 260.3 (13.0)                   & 0.9905 \\
      ThetaGPU & 67              & 0.9555          & 0.1178      & 238.5 (10.4)                   & 0.9901 \\
      \hline

      \multicolumn{6}{|c|}{Method 3: Random split on data points}                                          \\
      \hline
      HAL      & 71              & 0.95546         & 0.11760     & 420.4 (20.5)                   & 0.9901 \\
      DGX      & 150             & 0.9572          & 0.1131      & 236.9 (10.9)                   & 0.9908 \\
      ThetaGPU & 106             & 0.9563          & 0.1151      & 233.97 (10.8)                  & 0.9905 \\
      \hline
    \end{tabular}
  }
  \label{tab:diff_platform}
\end{table}

Table~\ref{tab:diff_platform} summarizes our key findings.
We can see that the area under the curve (AUC) of the receiver operating characteristic (ROC) curve the validation accuracy are stable at around 99 and 95.5\%, respectively.
These results are robust to data split methods, and agnostic to the underlying hardware used.

\subsubsection{Portability across software frameworks}
\label{sec:portability_software}
Here we explore the portability of the AI model across software frameworks that are extensively used for AI research, optimal assembly of software and hardware solutions, and containers.

\paragraph{ONNX and TensorRT conversion}

Software frameworks such as PyTorch and TensorFlow are extensively used in the AI community.
ONNX has emerged as a tool to ease the portability of models developed across software frameworks, and to optimize AI models for accelerated inference using tools such as NVIDIA TensorRT.
ONNX has also become a common standard to share and publish ML models.
Thus, we have quantified the performance of our IN model in three different implementations: PyTorch, TensorRT and ONNX.
The metrics used for this study are inference accuracy, running time, and AUC score.

We carried out these experiments on the ThetaGPU supercomputer using Python~3.6.3, ONNX~1.10.1, PyTorch~1.9.1, and TensorRT~8.2.1.8.
For inference, we considered a CMS test set consisting of 1800\,k test events/samples, and then quantified the performance and reliability of our three IN model implementations using the first 10\,k events in the test data.
We set the batch size to 1 for these comparisons.
The output of the IN model in these experiments is an array with two values that indicate the probability for the classification of two types of jets.
The results of these studies are summarized in \tablename~\ref{tab:diff_framework_table}.

\begin{table}[htbp]
  \centering
  \caption{
    Inference results, produced in the ThetaGPU supercomputer, for different frameworks using partial test data, all test data, and all test data within an Apptainer container.
  }
  \begin{tabular}{|c|c|c|c|}
    \hline
    Framework & Accuracy & Time/batch [s] & AUC                      \\
    \hline
    \multicolumn{4}{|c|}{Inference using partial test data}          \\
    \hline

    PyTorch   & 0.9735   & 0.0011         & 0.9915                   \\
    \hline
    ONNX      & 0.9735   & 0.0008         & 0.9915                   \\
    \hline
    TensorRT  & 0.9735   & 0.0099         & 0.9915                   \\
    \hline
    \multicolumn{4}{|c|}{Inference using all test data}              \\
    \hline

    PyTorch   & 0.97507  & 0.00123        & 0.99103                  \\
    \hline
    ONNX      & 0.97507  & 0.01264        & 0.99103                  \\
    \hline
    TensorRT  & 0.97507  & 0.01231        & 0.99103                  \\
    \hline
    \multicolumn{4}{|c|}{Inference on all test data using Apptainer} \\
    \hline
    PyTorch   & 0.97569  & 0.00098        & 0.99122                  \\\hline
    ONNX      & 0.97507  & 0.01224        & 0.99103                  \\
    \hline
    TensorRT  & 0.97506  & 0.01232        & 0.99103                  \\
    \hline
  \end{tabular}
  \label{tab:diff_framework_table}
\end{table}

We also tested these three different implementations using all 1800\,k events, using a batch size equal to 128.
The results of these two experiments are reported in \tablename~\ref{tab:diff_framework_table}.
Inference with the ONNX model is done on the GPU while the data are stored on the host side.
Thus, before inference, the data need to be copied from host to device.
The time/batch column refers to the time used to run one batch, including the data transfer between the two sides (device, host) and the inference part in the GPU device.
When we increase the batch size from 1 to 128, the running time becomes larger because the time taken to transfer the data increases.

For the second case, using a subset of the test set, we can see that when converting from PyTorch to TensorRT, the inference accuracy and AUC score are similar, and the running time of ONNX and TensorRT is shorter due to the accelerating effect of these two formats. 
When we used the entire test set, the running time of ONNX and TensorRT increase because we use a larger batch size.

\begin{figure}[htpb]
  \centering
  \includegraphics[width=0.7\textwidth]{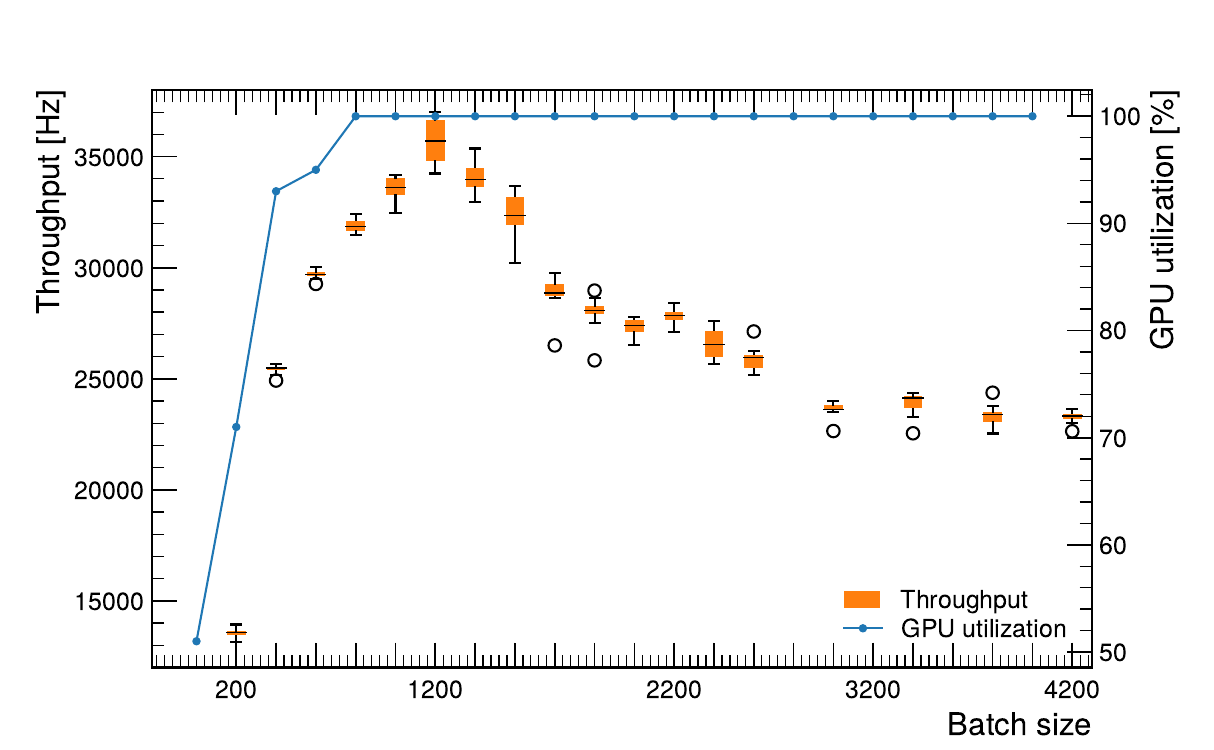}
  \caption{
    GPU utilization (shown as a blue line) and throughput (shown as box-and-whisker plots) as a function of batch size.
    GPU utilization saturates at 100\% for a batch size of 1000, while throughput peaks at 35\,k inferred events per second for a batch size of 1200.
    For the box-and-whisker throughput plots, ten runs are performed with a given batch size.
    The black line represents the median value of the throughput, the orange box represents the range from the first quartile to the third quartile, and the whiskers extend an additional distance of 1.5$\times$ the interquartile range.
    The white circles represent the outliers.
  }
  \label{fig:util-throughput}
\end{figure}

\paragraph{GPU utilization and throughput}
Since NVIDIA TensorRT was developed to optimize AI models for accelerated inference, we have quantified the interplay between batch size, GPU utilization, throughput, and inference accuracy.
In this context, throughput corresponds to the number of inferred events per second.
In practice, throughput is calculated by computing the total number of inferences divided by total time, or batch size divided by the average running time per batch.
Here, running time corresponds to the time taken to complete the analysis of one batch, including data transfer between device-host and the inference part at the GPU.
In our experiments, we increased the batch size from 100 to 2400 with a step size equal of 200, while from 2400 to 4200, we used a step size of 400.
For each batch size, we run 10 times and draw a boxplot of the throughput.
Our findings are summarized in Figure~\ref{fig:util-throughput}.
At a glance, we see that GPU utilization saturates at 100\% for a batch size of 1000, while throughput peaks (35\,k inferred events per second) at a batch size of 1200.
These findings exhibit the realm of applicability of TensorRT, i.e., for large scale ML inference workflows.

\subsection{Model interpretability}
\label{sec:interpretability}
In recent years, advances in explainable artificial intelligence (XAI)~\cite{MILLER20191} have made it possible to identify novel connections between an AI model's inputs, architecture, optimization, and predictions~\cite{xAI-intro,xAI-review,xAI-review-2}.
A substantial subset of XAI methods have been developed to analyze computer vision models where an intuitive reasoning can be extracted from human-annotated datasets to validate XAI techniques.
However, in other data structures, like large tabular data or relational data constructs like graphs, the use of XAI methods is still quite new~\cite{sahakyan2021explainable, xAI-GNN-survey}.
These XAI techniques have been harnessed across disciplines to quantify the reliability of AI models for science~\cite{zhang2018visual,asad:2018K,viztsne2,khan_huerta_zheng_forecast}.
Recently, the scope of XAI has been expanded to include AI applications within HEP~\cite{neubauer2022explainable,shanahan2022snowmass,pmlr-v162-miao22a,miao2023interpretable}.
In HEP, XAI has been used to understand the output of AI models used in high energy detectors~\cite{9302535}, including parton showers at the LHC~\cite{LAI2022137055}, deep-neural-network-based classification of jets~\cite{Agarwal:2020fpt,khot2022detailed}, and particle-based global event description algorithms~\cite{LRP-MLPFlow}.
Learnable randomness injection (LRI)~\cite{miao2023interpretable} provides interpretability by identifying a subset of HEP detector hits in a particle cloud that is the most relevant to the prediction results.
This method can also identify whether the existence or specific geometry of a point is important.

\subsubsection{Evaluating feature importance}
\label{sec:feat}

Identifying feature importance has been a significant component of XAI methods and has been thoroughly studied in the context of classification models~\cite{tang2014feature}.
In standard feature selection tasks, a reasonable subset of the features that excels in some model performance metric is chosen.
Although it is conceptually different from feature ranking in \textit{post-hoc} model interpretation, the latter usually also relies on minimizing a model's performance loss~\cite{ribeiro2016should}.
One of the most useful model analysis tool of a binary classification is the ROC curve, and the corresponding area under the curve (AUC) serves as a scalar metric for evaluating model performance.
AUC-based feature ranking has been widely used in the AI literature~\cite{chen2008fast,wang2009feature,serrano2010feature}.
We adapt those same principles for our model interpretation studies.
One strategy for evaluating a feature's contribution in making predictions is to investigate the model's performance when that feature is \emph{masked}, e.g., by replacing it with a population-wide average value or a zero value, whichever is contextually relevant to the model's relationship with the training dataset.

\begin{figure}[!t]
  \centering
  \includegraphics[width=0.8\textwidth]{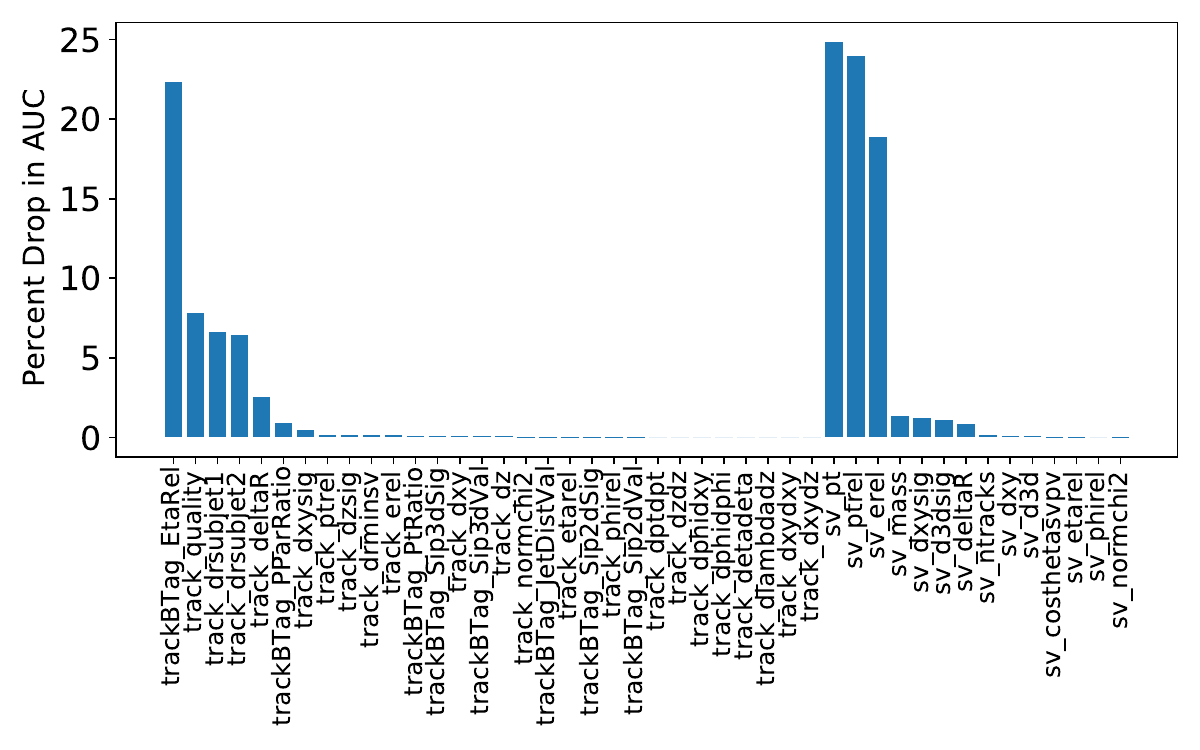}
  \caption{
    Change in AUC score with respect to a baseline model when each of the tracks and secondary vertex (SV) features are individually masked during inference.
  }
  \label{fig:dAUC}
\end{figure}

In order to identify the features that play the most important role in the IN model's decision-making process, we first train the model with its default settings, which we call the \textit{baseline} model.
During the training, for any event where certain input tracks or secondary vertices are absent for a given jet, its corresponding entries are marked with zeros.
Hence, we mask one feature at a time for all input tracks or secondary vertices by replacing the corresponding entries by zero values.
We obtain predictions from the trained model and evaluate the AUC score.
The change observed in the AUC score when masking each of the features is presented in Figure~\ref{fig:dAUC}.
It shows that while the model has been trained to take into account the entire feature space, there are 14 track features and 4 secondary vertices features that, if removed one at a time, reduce the model's AUC score by less than 0.05\%.
Inspired by computer vision studies, we propose that the input features that cause the largest change in the AUC score may be regarded as the features that play the most important role in the model's decision-making process.

The weak dependence of the model on many of its input features indicates that the model can learn the jet classification task from a subset of input features.
To further investigate the cumulative impact of removing these \textit{unimportant} features, we mask multiple features at the same time based on a few arbitrary thresholds for the change in AUC score compared to the baseline.
The set of masked features includes every track and secondary vertex  feature that causes a change in AUC score below that threshold when independently masked.
To compare how individual predictions vary on average, we compute the model fidelity score~\cite{fidelity,xAI-GNN-survey}, defined as
\begin{equation}
  F(\mathcal{M}_1, \mathcal{M}_2) = 1 - \frac{1}{N}\sum_{i=0}^{N-1} \left|\hat{y}_i^1 -  \hat{y}_i^2\right|~\,.
  \label{eq:fidelity}
\end{equation}

Here, $\mathcal{M}_1$ and $\mathcal{M}_2$ are two different models and the corresponding classifier scores for the $i$-th data sample are respectively given by $\hat{y}_i^1$ and  $\hat{y}_i^2$.
The results are summarized in Table~\ref{tab:dAUC}.
The model's performance, both in terms of AUC and fidelity scores, remains very close to the baseline even when masking up to 14 particle track and 4 secondary vertex features.

\begin{table}[htbp]
  \centering
  \caption{
    The table shows the performance of a baseline model when multiple features are simultaneously masked based on AUC score drop threshold.
    $\Delta P$ ($\Delta S$) represents the number of particle (secondary vertices) features that have been masked.
    The fidelity score, see Equation~(\ref{eq:fidelity}), is measured with respect to the baseline model.
  }
  \begin{tabular}[b]{ |c|c|c|c|c| }
    \hline
    Threshold [\%] & $\Delta P$ & $\Delta S$ & AUC [\%] & Fidelity [\%] \\
    \hline
    0              & 0          & 0          & 99.02    & 100           \\
    0.001          & 8          & 2          & 99.02    & 99.69         \\
    0.005          & 9          & 2          & 99.05    & 99.46         \\
    0.01           & 11         & 3          & 98.99    & 98.85         \\
    0.05           & 14         & 4          & 98.81    & 97.12         \\
    1.00           & 25         & 8          & 80.49    & 70.83         \\
    \hline
  \end{tabular}
  \label{tab:dAUC}
\end{table}

While the AUC and fidelity scores allow determining which features play important roles in the IN's decision making process, we can inspect the importance of these features for individual tracks and vertices by the layerwise relevance propagation (LRP) technique~\cite{LRP-NN, LRP-overview}.
The LRP technique propagates the classification score predicted by the network backwards through the layers of the network and attributes a partial relevance score to each input.
The original LRP method has been developed for simple MLP networks.
Variants of this method have been explored to propagate relevance across convolutional neural networks~\cite{LRP-pixel,Agarwal:2020fpt} and graph neural networks~\cite{schnake2021higher, LRP-MLPFlow}.

Since some of the input features show a high degree of correlation with each other, we use the LRP-$\gamma$ method described by Montavon et al.~\cite{LRP-overview}, which is designed to skew the LRP score distributions to nodes with positive weights in the network and thus, avoiding propagation of large but mutually canceling relevance scores.
In order to apply the LRP method for the IN model, we propagate scores across (i) the aggregation of internal representation of track features obtained from the aggregator network
\begin{eqnarray}
  O_{[D_O \times N_p]} & \to &\bar{O}_{[D_O]}~\,,
\end{eqnarray}
and (ii) the interaction matrices that send edge-level representations to the individual particle tracks
\begin{eqnarray}
  {E}_{pp[D_E \times N_{pp}]} & \to &\bar{E}_{pp[D_E \times N_{p}]} \\ {E}_{vp[D_E \times N_{vp}]} &\to &\bar{E}_{vp[D_E \times N_{p}]}~\,.
\end{eqnarray}
The relevance scores for the output, $O_{[D_O \times N_p]}$, of the $f_O$ function can be obtained as
\begin{equation}
  r_{kn} = \bar{r}_k\left(\frac{o_{kn}}{\displaystyle\sum_m o_{km}}\right)\,,
\end{equation}
where $\bar{r}$ represents the LRP scores for the summed internal representation.
On the other hand, the relevance scores $\bar{R}_{kn}$ for the track level internal representations in $\bar{E}_{pp[D_E \times N_{p}]}$ can be propagated to edge level representations, $E_{pp[D_E \times N_{pp}]}$, using the relation
\begin{equation}
  r_{km} = e_{km}\sum_n \left(\frac{\bar{r}_{kn}}{\bar{e}_{kn}}\right) \left(R_{R}\right)_{nm}~\,,
\end{equation}
where $R_{R}$ is the receiver matrix for particle-particle interactions.
A similar expression allows translating the relevance scores of the track level representation in $\bar{E}_{pv[D_E \times N_{p}]}$ to  track-vertex edge representations in ${E}_{vp[D_E \times N_{vp}]}$ using the the receiver matrix $R_K$ for vertex-particle interactions.

\begin{figure}[!htbp]
  \centering
  \includegraphics[width=0.95\textwidth]{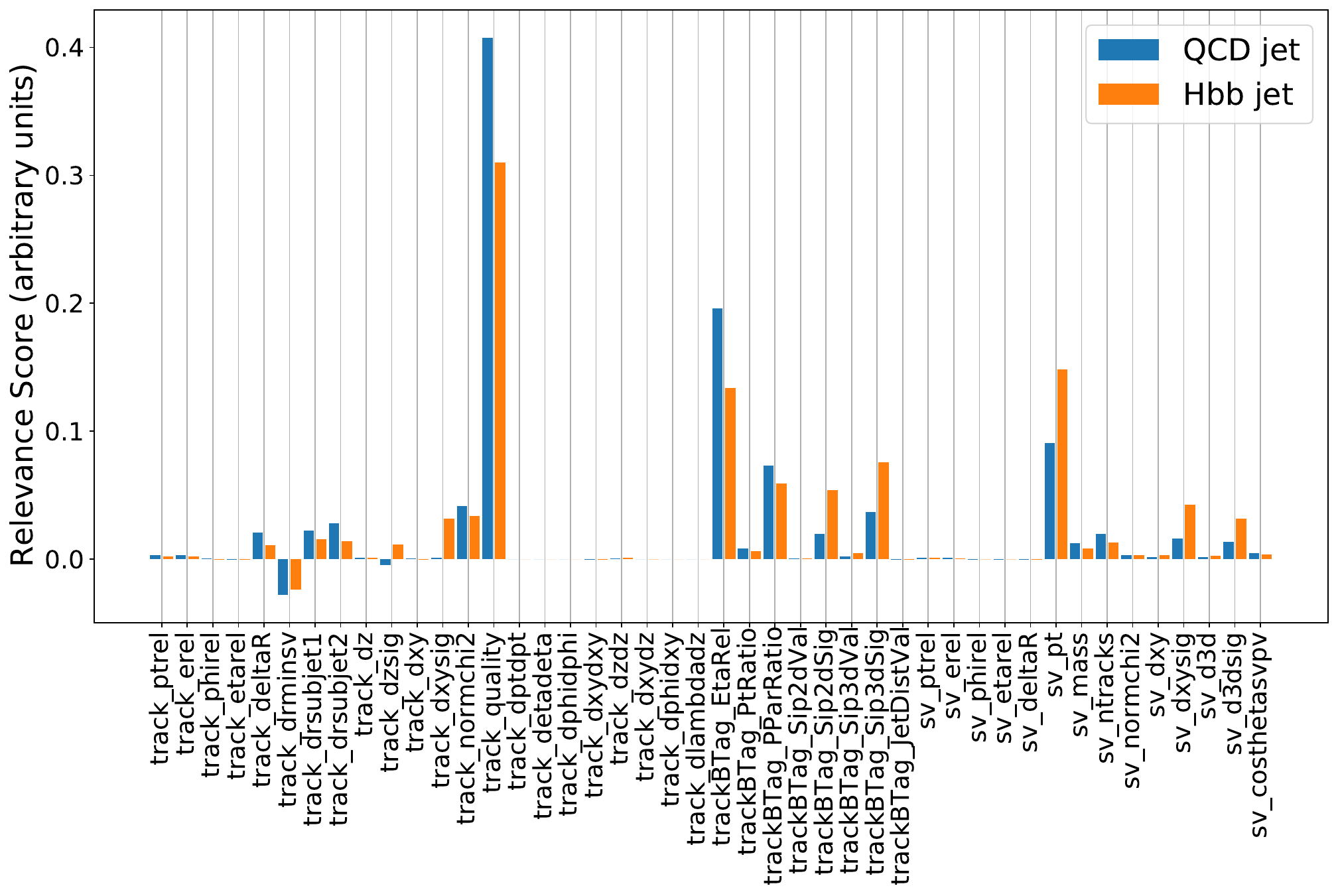}\\
  \includegraphics[width=0.93\textwidth]{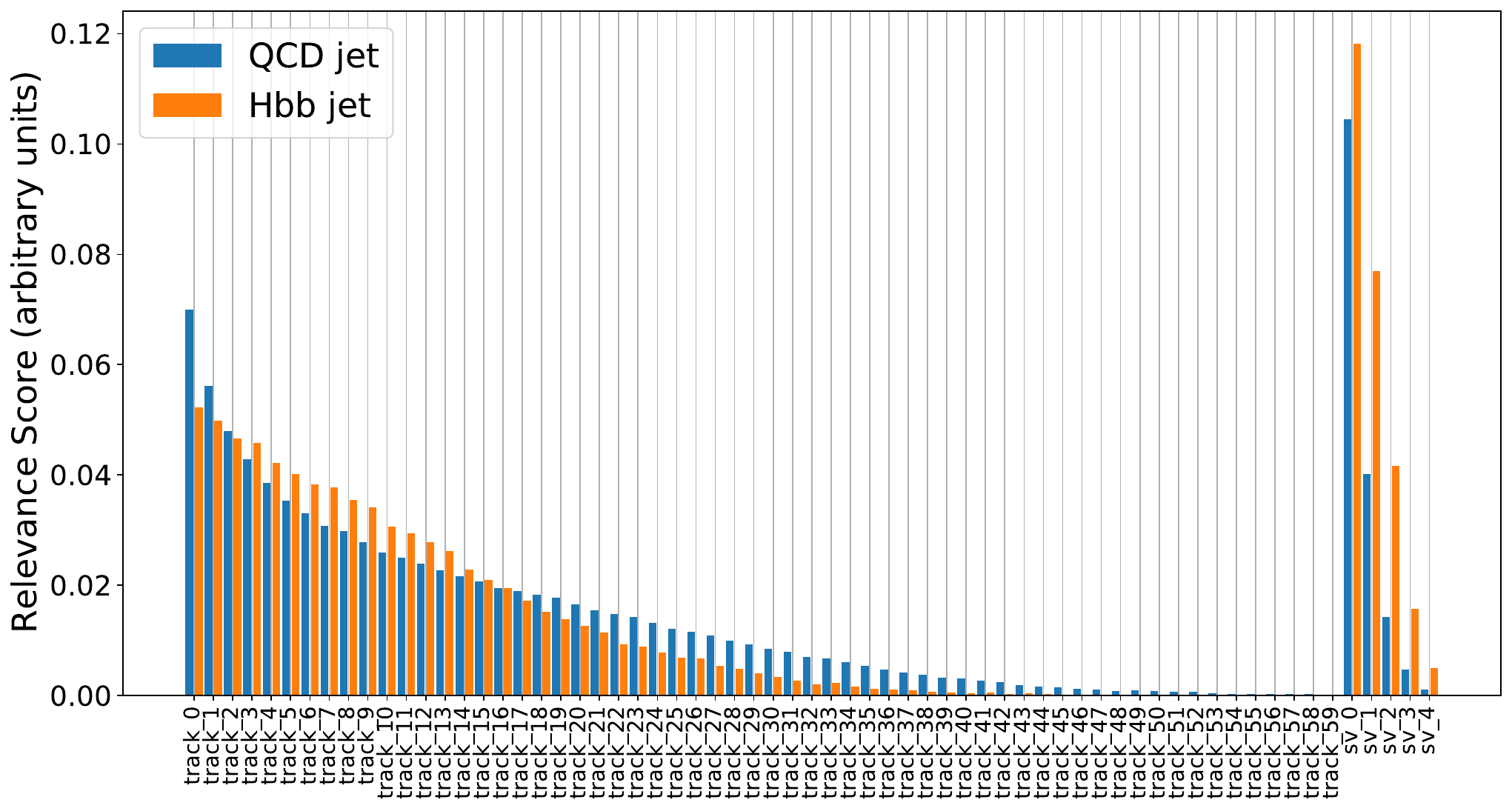}
  \caption{Average relevance scores attributed to input track and secondary vertex features (upper) and individual tracks and secondary vertices (lower).
    The tracks and secondary vertices (SVs) are ordered according to their relative energy with respect to the jet energy.}
  \label{fig:LRP}
\end{figure}

We show the average scores attributed to the different features for QCD and \Hbb jets in Figure~\ref{fig:LRP}.
When compared with the change in AUC score by individual features in Figure~\ref{fig:dAUC}, the track and secondary vertex features with largest relevance scores are also the features that individually cause the largest drop in AUC score.
We additionally observe that the track features are generally assigned larger relevance scores for QCD jets and secondary vertex features play a more important role in identifying the \Hbb jets.
This behavior is also justified from a physics standpoint, since the presence of high energy secondary vertices is an important signature for jets from \PQb quarks because of its relatively longer lifetime.
This is also illustrated in Figure~\ref{fig:LRP}, where the cumulative relevance score for each track and vertex is shown.
The tracks and vertices are ordered according to their relative energy and our results show that the higher energy tracks and vertices are generally attributed with higher relevance scores for both jet classes.
However, feature representing relative track energy, \texttt{track\_erel}, itself does not carry notable relevance weight.
On the other hand, the relevance attributed to \texttt{sv\_pt}, which is strongly correlated with \texttt{sv\_erel}, is very large.

\begin{figure}[!htbp]
  \centering
  \includegraphics[width=0.32\textwidth]{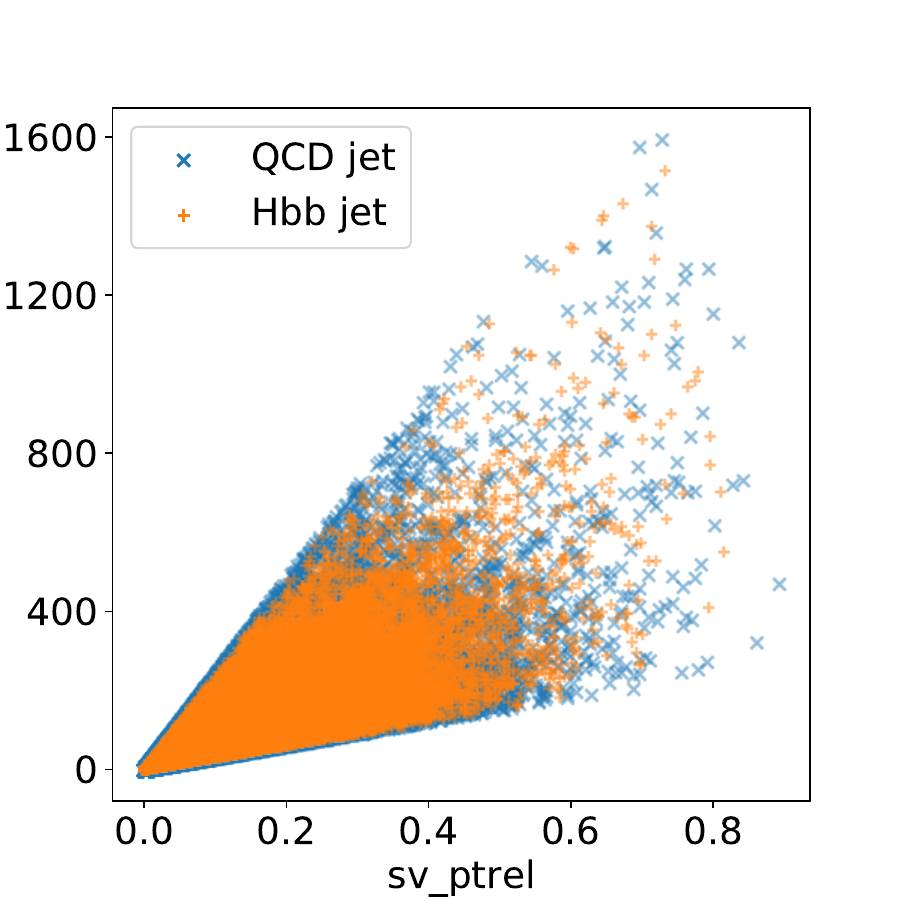}
  \includegraphics[width=0.32\textwidth]{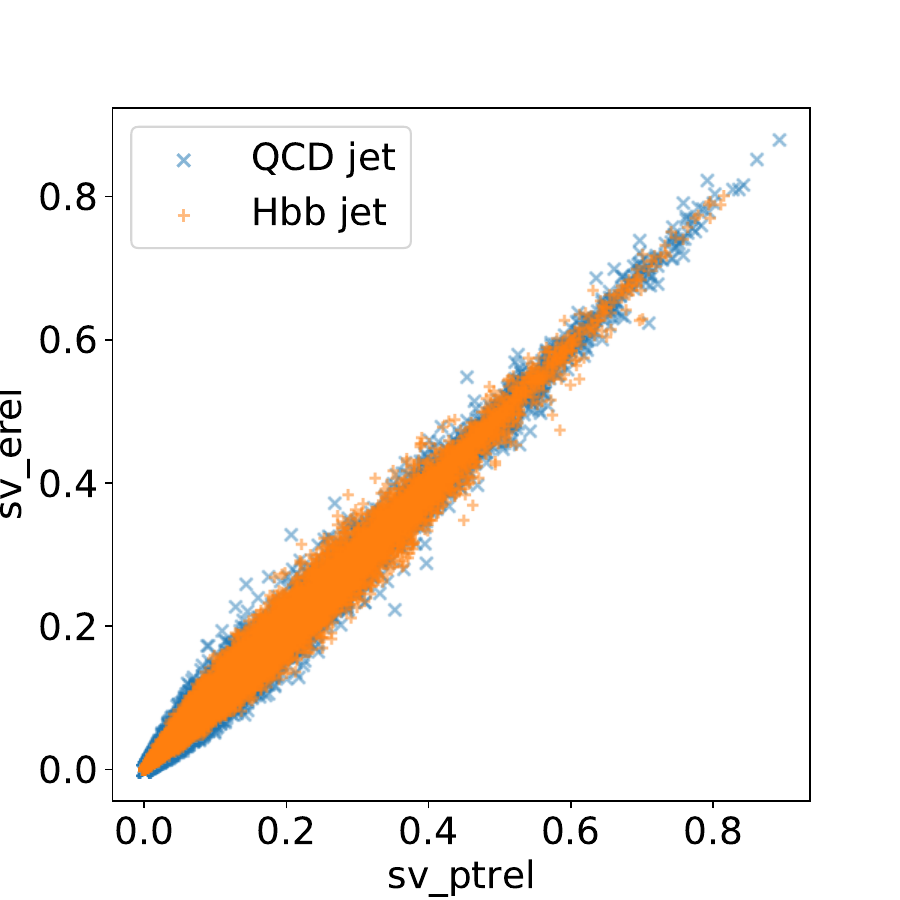}
  \includegraphics[width=0.32\textwidth]{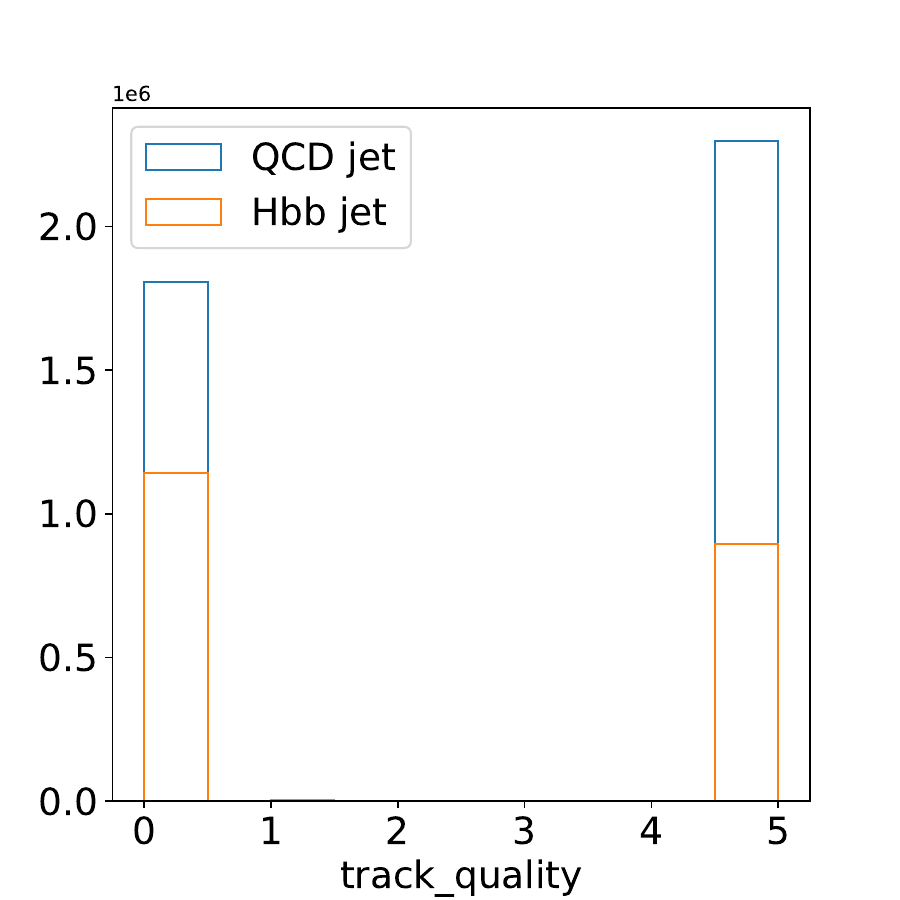}
  \caption{
    Scatter plots of  \texttt{sv\_ptrel} and \texttt{sv\_pt} (left) \texttt{sv\_ptrel} and \texttt{sv\_erel} (middle), and distribution of the categorical variable \texttt{track\_quality} (right). \texttt{sv\_ptrel} and \texttt{sv\_erel} represent the relative transverse momentum and energy of the secondary vertex with respect those of the jet.
    \texttt{sv\_pt} is the transverse momentum of the secondary vertex. \texttt{track\_quality} is a categorical variable to represent the quality of track reconstruction where the peak at 0 represents absent tracks.
  }
  \label{fig:scatter}
\end{figure}

We also note that while the secondary vertex features \texttt{sv\_ptrel} and \texttt{sv\_erel} are assigned relatively low relevance scores, masking them independently leads to very large drops in the AUC score.
This \textit{apparent} discrepancy can be explained by the very high correlation between these variables, each of which also displays a very large correlation (correlation coefficient of 0.85) with \texttt{sv\_pt}, as shown in Figure~\ref{fig:scatter}.
Because the LRP-$\gamma$ method skews the relevance distribution between highly correlated features, it suppresses the LRP scores for those two variables while assigning a large relevance score to the variable \texttt{sv\_pt}.

We make an additional observation regarding the importance attributed to the feature called \texttt{track\_quality}.
This feature is a qualitative tag denoting the track reconstruction status, and has an almost identical, doubly peaked distribution for both jet categories.
In Figure~\ref{fig:scatter}, the peak at 0 represents absent tracks.
With such an underlying distribution, this variable does not contribute to the classifier's ability to distinguish the jet categories.
However, the large relevance score associated with it, along with the large drop in AUC score upon masking this feature, indicates that the classifier's class-predictive output for each class somehow receives a large contribution from the numerical embedding used to represent this feature and eventually gets canceled by the softmax operation.

We have found that the two previously mentioned secondary vertex features, along with \texttt{track\_quality}, have no discernible impact on the IN model's ability to tell the jet categories apart by retraining the model without these variables.
The model that was trained without these variables, along with the 11 (3) track (secondary vertex) features that report a change in AUC of less than 0.01\%, converged with an AUC score of 99.00\%.
In the absence of these redundant features, we observed some differences in the relative distribution of the relevance scores.
Thus, we are better able to understand which features play a more important role in the identification of \Hbb or QCD jets, respectively.
These physics-informed validation of model explanation pinpoints two major drawbacks of the existing XAI methods.
First, explanations for models trained with highly correlated input features can be inconsistent across approaches and second, treating categorical and continuous variables on equal footing in XAI methods might lead to misleading attribution of feature importance.

\paragraph{Inspecting the activation layers}
\label{sec:NAP}

Here we aim to gain new insights on the IN model's decision-making process at the layer level.
As the IN processes the input, it is passed through three different MLPs that approximate arbitrary nonlinear functions identified as $f_R^{pp}, f_R^{vp},$ and $f_O$.
In order to explore the activity of each neuron and compare it with the activity of neurons in the same layer, we define relative neural activity (RNA)~\cite{khot2022detailed} as
\begin{equation}
  \mathrm{RNA}(j,k;\mathcal{S}) = \frac{\displaystyle\sum_{i=1}^{N} a_{j,k} (s_i)}{\displaystyle\max_j\sum_{i=1}^{N} a_{j,k}(s_i)}
\end{equation}
where $\mathcal{S} =\{s_1, s_2, \dots, s_N\}$ represents a set of samples over which the RNA score is evaluated.
The quantity $a_{j,k} (s_i)$ is the activation of $j$-th neuron in the $k$-th layer when the input to the network is $s_i$.
When summed over all the samples in the evaluation set $\mathcal{S}$, this represents the cumulative neural response of a node, which is normalized with respect to the largest cumulative neural response in the same layer to obtain the RNA score.
Hence, in each layer, there will be at least one node with an RNA score of 1.
Since the neurons are activated with ReLU activation in the IN model, the RNA score will be strictly between 0 and 1.

At a qualitative level, this study aims to identify which neurons are most actively engaged when the IN model produces an output.
Since the MLPs in the IN model consist of only fully-connected layers, each layer takes all the activations from the previous layer as inputs.
As all nodes within a given layer are subject to the same set of inputs, we can reliably estimate how strongly they perceive and transfer that information to the next layer by looking at their activation values.
For the same reason, we normalize the cumulative activation of a node with respect to the largest aggregate in the same layer.

\begin{figure}[!htbp]
  \centerline{
    \includegraphics[width=0.52\textwidth]{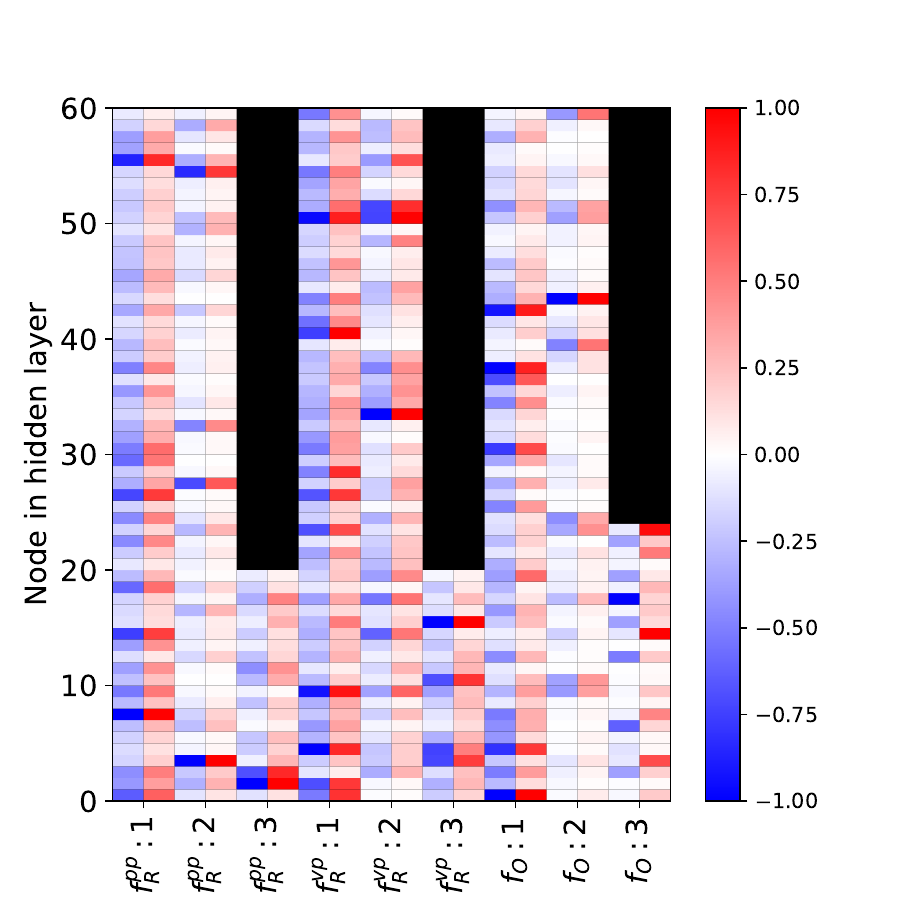}
    \includegraphics[width=0.52\textwidth]{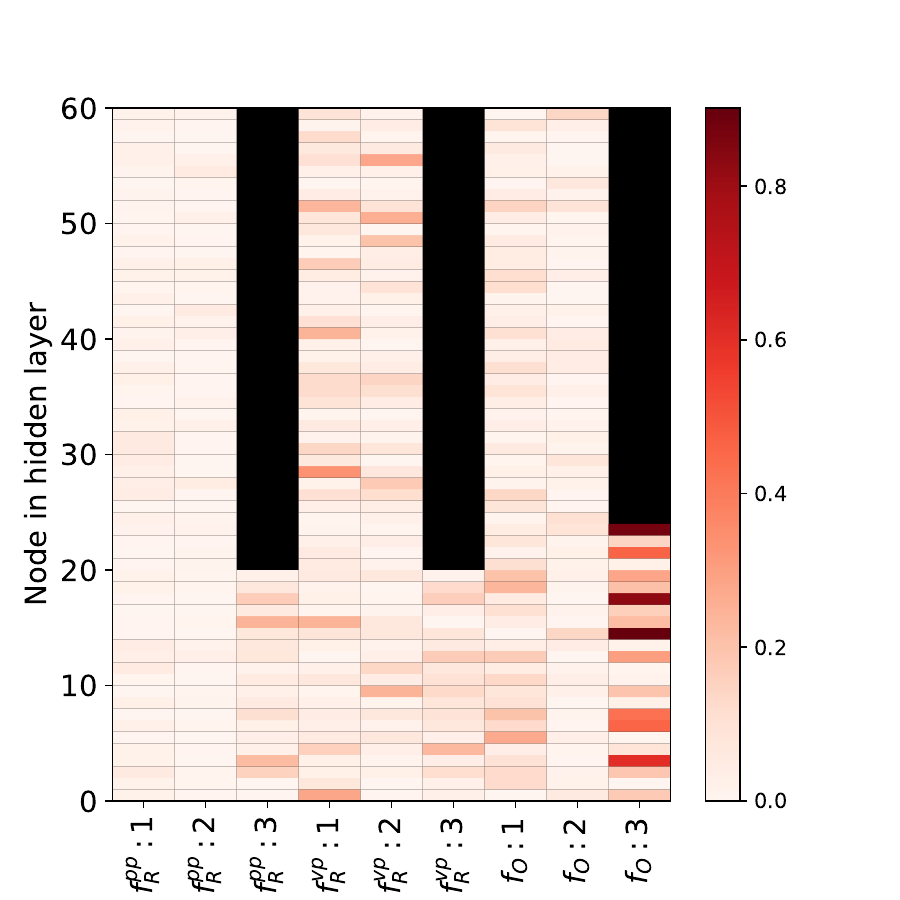}
  }
  \caption{
    2D map of relative neural activity (RNA) score
    for different nodes of the activation layers (left).
    To simultaneously visualize the scores for QCD and \Hbb jets, we project the RNA scores of the former as negative values.
    2D map of absolute difference in RNA score for QCD and \Hbb jets for different nodes of the activation layers (right).
    In both figures, the labels associated with the horizontal axis entries represent the nonlinear function and the layer associated with it.
  }
  \label{fig:NAPs-baseline}
\end{figure}

Figure~\ref{fig:NAPs-baseline} (left) shows the (NAP) diagram for the baseline model, showing the RNA scores for the different activation layers.
The scores are separately evaluated for QCD and \Hbb.
To simultaneously visualize these scores, we project the RNA scores of the former as negative values.
The NAP diagram clearly shows that the network's activity level is quite sparse.
In some layers, more than half of the nodes show RNA scores less than 0.2.
This implies that while some nodes are playing very important roles in propagating the necessary information, other nodes do not participate as much.
We additionally observe that right until the very last layer of the aggregator network $f_O$, the same nodes show the largest activity level for both jet categories.
This is better illustrated in Figure~\ref{fig:NAPs-baseline} (right), where the absolute difference in RNA scores for the two jet categories are mapped.
For most nodes in every layer but the very last one, the difference in RNA scores is very close to zero.
However, different nodes are activated in the last layer for the two jet categories, indicating an effective disentanglement of the jet category information in this layer.
However, even in this layer, the activity level appears to be sparse---only a few nodes showing large activation for each category.

\subsubsection{Model reoptimization}
The studies presented in Sections~\ref{sec:feat} and \ref{sec:NAP} suggest that the baseline IN model can be made simpler by reducing both the number of input features it relies on and the number of trainable parameters.
To explore this observation, we trained alternate variants of the IN models where the features \texttt{sv\_ptrel}, \texttt{sv\_etrel}, and \texttt{track\_quality} were dropped along with additional 11 track and 3 secondary vertex features that reduce the AUC less than 0.01\%, as shown in Figure~\ref{fig:dAUC}.

\begin{table}[htbp]
  \caption{
    The performance of a baseline and ablated models.
    $\Delta P$ represents the number of particle track features that have been dropped and $h$ is the number of nodes in the hidden layers.
    The fidelity score is measured with respect to a baseline model.
    Sparsity is measured by the fraction of activation nodes with an RNA score less than 0.2
  }
  \centering
  \begin{tabular}[b]{ |c|c|c|c|c|c| }
    \hline
    $\Delta P$, $\Delta S$ & $h, D_E, D_O$ & Parameters & AUC score [\%] & Fidelity [\%] & Sparsity \\
    \hline
    0, 0 (baseline)        & 60, 20, 24    & 25554      & 99.02          & 100           & 0.56     \\
    \hline
                           & 32, 16, 16    & 8498       & 98.87          & 96.93         & 0.52     \\
    12, 5                  & 32, 8, 8      & 7178       & 98.84          & 96.79         & 0.48     \\
                           & 16, 8, 8      & 2842       & 98.62          & 96.12         & 0.40     \\
    \hline
  \end{tabular}
  \label{tab:reopt-models}
\end{table}

The details and performance metrics of these models are given in \tablename~\ref{tab:reopt-models}.
It should be noted that the ablated models presented here  represent neither an exhaustive list of such choices nor any result of some rigorous optimization.
These results demonstrate that a simpler IN model may be developed without compromising the quality of its performance.
As can be seen from the results in \tablename~\ref{tab:reopt-models}, both AUC score and fidelity of the alternate models are very close to that of the baseline model, though the number of trainable parameters is significantly lower.

\begin{figure}[!htbp]
  \centerline{
    \includegraphics[width=0.52\textwidth]{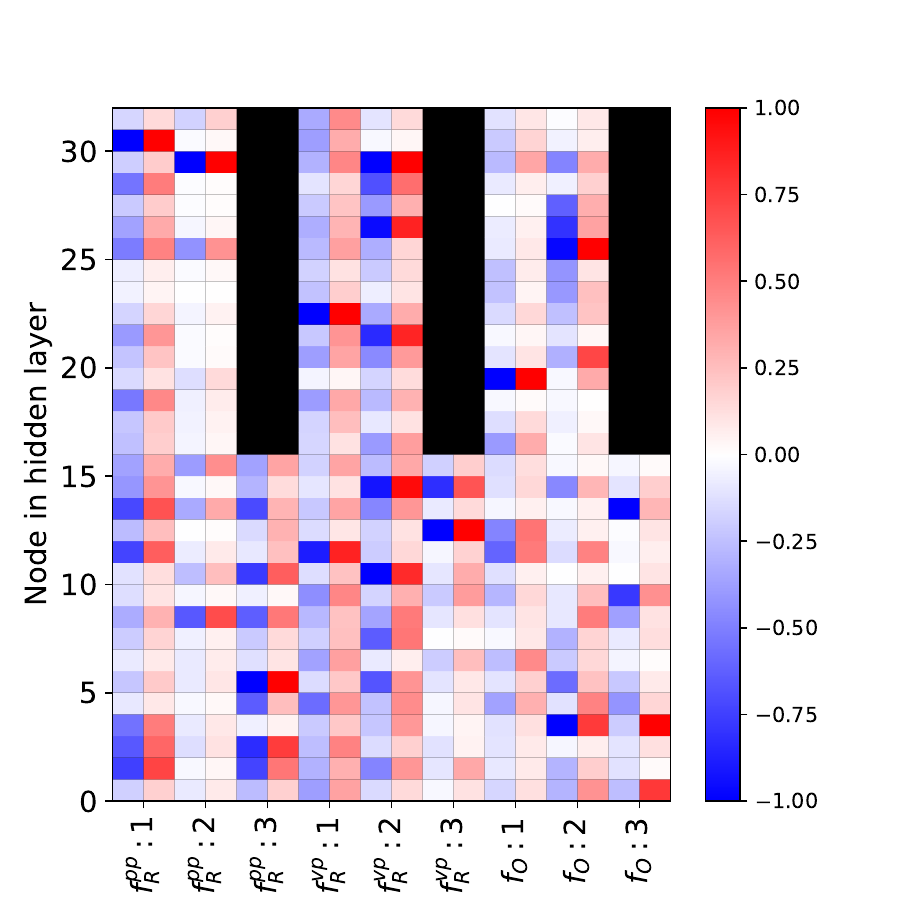}
    \includegraphics[width=0.52\textwidth]{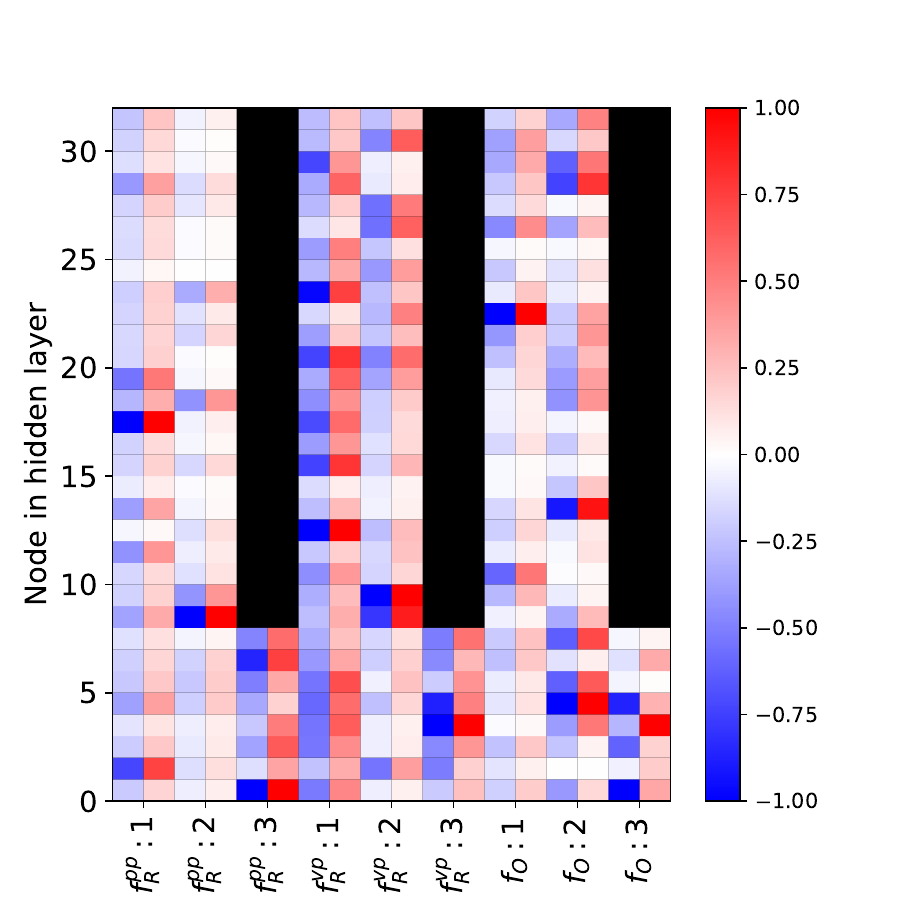}
  }
  \caption{
    Neural activation pattern diagrams for the IN model where the features \texttt{sv\_ptrel}, \texttt{sv\_erel}, \texttt{track\_quality} along with the additional 11 (3) particle track (secondary vertex) features associated with a change in AUC of 0.01\%.
    In both models, the number of nodes in hidden layers is 32 while $D_E = D_O = 16$ (left) or $D_E = D_O = 8$ (right).
  }
  \label{fig:NAPs-alt}
\end{figure}

Figure~\ref{fig:NAPs-alt} shows the NAP diagrams for the model with 15 (5) dropped track (vertex) features with 32 nodes per hidden layer where the internal representation dimensions $D_E$ and $D_O$ are set to 16 and 8 for the left and right figures, respectively.
Sparsity of the latter, as measured by the number of activation nodes with $\mathrm{RNA} < 0.2$, is noticeably lower than the baseline model though the former has increased sparsity.
With reduced size for the post interaction internal space representation, the alternate models do not completely disentangle the jet classes at the output stage of $f_O$.

\section{Discussion and conclusion}
\label{sec:summary}
We have proposed a practical definition of findable, accessible, interoperable, and reusable (FAIR) principles for machine learning (ML) and artificial intelligence (AI) models in experimental high energy physics (HEP).
To promote adherence to these principles, we have introduced a FAIR AI project template and demonstrated how to implement this template with a model to identify Higgs bosons decaying to bottom quarks.
We studied the robustness of this FAIR AI model and its portability across hardware architectures and software frameworks, and reported new insights on the interpretability of AI predictions, by studying the interplay between FAIR datasets and AI models.

These studies represent a step towards a FAIR ecosystem of data and AI models to enable and streamline automated AI-driven scientific discovery across disciplines~\cite{huerta2022fair}.
Future work in this area will need to address many outstanding issues, such as providing documentation in a machine-readable way, as well as the development of standardized application programming interfaces (APIs) for federating searching, accessing, and interoperating AI models hosted on different platforms, such as GitHub, DLHub, AI Model Share, and HuggingFace.
We also stress that the FAIR principles outlined in this paper are by no means an exhaustive prescription for shareable, reproducible, and extendable scientific AI research.
Nonetheless, we recommend the adoption of this FAIR AI model standard to advance HEP research.

\section*{Acknowledgments}
This research is supported by DE-SC0021258, DE-SC0021395, DE-SC0021225, and DE-SC0021396 from the Office of Advanced Scientific Computing Research (ASCR) within U.S. Department of Energy (DOE) Office of Science, by the FAIR Data Program of the DOE, Office of Science, ASCR, under contract number DE-AC02-06CH11357, and by Laboratory Directed Research and Development (LDRD) funding from Argonne National Laboratory, provided by the Director, Office of Science, of the DOE under Contract No. DE-AC02-06CH11357.
It used resources of the Argonne Leadership Computing Facility, which is a DOE Office of Science User Facility supported under Contract DE-AC02-06CH11357, and resources supported by the U.S. National Science Foundation's Major Research Instrumentation program, grant \#1725729, as well as the University of Illinois at Urbana-Champaign.
We thank Nikil Ravi, Pranshu Chaturvedi, and Huihuo Zheng for expert support creating and deploying ONNX and TensorRT engines, and Apptainer containers in the ThetaGPU supercomputer.

\section*{Author contributions statement}
JD conceptualized some of the original Cookiecutter template ideas, supervised, and organized the effort.
IHK and HL developed the \texttt{cookiecutter4fair} project template.
DSK provided expertise on FAIR for other types of objects, such as software, and other disciplines, such as computer science.
RZ and AR studied the portability and interpretability of the interaction network model and VVK advised on computing platforms and tools used in this study.
MSN advised on the use of ONNX for portability and interpretability of the interaction network model.
EAH guided activities on the use of DLHub and the deployment and use of ML models on disparate high performance computing platforms.
All authors contributed to writing and reviewing the work and the manuscript.

\section*{Competing interests}
The authors declare no competing interests.

\bibliographystyle{cms_unsrt}
\bibliography{sample}

\end{document}